\documentclass[smallextended]{svjour3}       

\smartqed  

\usepackage{graphicx}
\usepackage{color}
\usepackage{graphics,graphicx,epsfig,subfigure}
\usepackage{latexsym}
\usepackage{amssymb}
\usepackage{mathrsfs}
\usepackage{amsmath}
\usepackage{float}
\usepackage{psfrag}
\usepackage{rotating}
\usepackage{longtable}
\usepackage{supertabular}
\usepackage[misc]{ifsym}
\usepackage{gensymb}
\usepackage[round]{natbib}
\usepackage{hyperref,xcolor}

\begin{document}

\title{An improved Probabilistic Seismic Hazard Assessment of Tripura, India}

\author{Suman Sinha
\and
       S. Selvan
}

\institute{Suman Sinha \Letter \and S. Selvan \at
Engineering Seismology Division,
Central Water and Power Research Station, Pune, India\\
\email {suman.sinha.phys@gmail.com}
}

\maketitle

\begin{abstract}
The State of Tripura lies in northeast India which is
considered to be one of the most seismically active 
regions of the world. In the present study,
a realistic Probabilistic Seismic Hazard 
Assessment (PSHA) of Tripura State based on
improved seismogenic sources considering layered 
polygonal sources corresponding 
to hypo-central depth range of $0-25$ km, 
$25-70$ km and $70-180$ km, respectively and data driven
selection of suitable Ground Motion Prediction 
Equations (GMPEs) in a logic
tree framework is presented. Analyses have been carried out by 
formulating a layered seismogenic source zonation
together with smooth-gridded seismicity. Using the
limited accelerogram records available, most suitable
GMPEs have been selected after performing a thorough quantitative
assessment and thus the uncertainty in selecting 
appropriate GMPEs in PSHA has been addressed by combining 
them with proper weight factor. The computations of seismic
hazard are carried out in a higher resolution of grid
interval of 0.05$\degree$ $\times$ 0.05$\degree$.
The probabilistic seismic hazard 
distribution in terms of Peak 
Ground Acceleration (PGA) and $5\%$ damped Pseudo 
Spectral Acceleration (PSA) at 
different time periods for $10\%$ and $2\%$
probability of exceedance in $50$ years at
engineering bedrock level have been presented.
The final results show significant improvements
over the previous studies, which is reflecetd in
the local variation of the hazard maps.
The design response spectra at engineering 
bedrock level can be computed
for any location in the study region from the hazard
distributions. The results will be useful for
earthquake resistant design and construction of 
structures in this region.
\keywords{Layered polygonal seismogenic source zones 
\and Probabilistic seismic hazard 
\and Smooth gridded seismicity 
\and Ground motion prediction equation
\and Ranking of GMPEs}
\end{abstract}

\section{Introduction}
\label{intro}
Hazards associated with earthquakes are commonly referred to as 
seismic hazards and ground shaking is considered to be the most important
of all seismic hazards because all the other hazards are consequences of 
ground shaking \citep{kra1}. Ground shaking causes extensive damage to properties and
lives in a seismic prone terrain possessing favourable seismo-tectonics and
local geological site conditions. It thus necessitates to estimate seismic 
hazard of the terrain under consideration in a realistic way, which could 
provide the necessary design inputs for earthquake resistant design of 
structures.

Northeast India is one of the most highly seismic active regions which 
record more than seven earthquakes of magnitudes 5 and above per year on
an average \citep{stg1}. The State of Tripura is one
of the eight northeastern states of India and from the seismological point 
of view, Tripura is considered to be of interest due to the existence of the
important geotectonic unit Tripura Fold Belt to the west of Indo-Burmese 
ranges. Compressive movements between the Indian plate and the Eurasian plate
during Oligocene to recent times produced the Tripura Fold Belt \citep{gupta2}. 
Besides that, Tripura is surrounded by
Eastern Himalaya and the Shillong
Plateau-Mikir Hills in the north, the Naga-Disang Thrust
system in the northeast, the Tertiary fold-thrust belt
of the Indo-Burma Arc in the east and the Bengal Basin in
the southwest.

In the region surrounded by Tripura, 18 events of magnitude $M_w > 7.0$ 
have taken place since 1664 (shown in Figure \ref{ros})
and the impacts of these earthquakes have attracted the attention of scientists
and engineers towards seismic safety evaluation of the region. The latest 
seismic zoning map of India released by Bureau of Indian Standards (BIS) 
assigns four levels of seismicity in terms of zone factor \citep{bis1}, which is
twice the zero period acceleration (ZPA) of the design spectrum. The present
study area falls in seismic zone $V$ with the highest zone factor $0.36$ as
per \citet{bis1}. However, the crucial limitation of the 
seismic zonation code of India \citep{bis1} is that it is not based on 
comprehensive seismic hazard analysis and hence it lacks the probabilistic
features \citep{khat1}. However, the probabilistic approach in the computation
of seismic hazard analysis is considered to be more appropriate and realistic 
for its scientifically sound background \cite{kra1}.  PSHA approach has the property
that with a specified confidence level, the ground motion will not be exceeded
at any of the time periods due to any of the earthquakes expected during a given
time interval, thus taking the randomness of the earthquake occurrences in space,
time and magnitude into account.
The computational formulation of probabilistic seismic hazard 
assessment (PSHA) was developed by \citet{corn2,mcg1}.

\citet{basu1} and \citet{khat2} adopted a probabilistic approach to 
prepare seismic 
zonation maps in terms of peak ground acceleration (PGA) for a specific 
return period. \citet{bha1} performed a probabilistic seismic hazard assessment 
(PSHA) of India under the Global Seismic Hazard Assessment Program (GSHAP). 
Many researchers have also carried out the seismic hazard studies of 
northeastern states of India in the past \citep{shar2,nath5,thin1,
ragh2,stg1,das2}. 
The present study attempts
to improve upon the existing studies to some extent by considering layered 
polygonal sources corresponding to different 
hypo-central depths together with smooth-gridded seismicity and data driven 
selection of ground motion prediction equations (GMPEs). Similar type of 
methodology has been adopted previously by \citet{nath2,nath4}. 
However, the computation
of seismic hazard in the present study is carried out in a higher resolution
of finer grid interval of $0.05 \degree \times 0.05 \degree$ and the 
suitability of different GMPEs against recorded strong motion data of
different focal depths are judged by the histograms of normalized residuals
and the likelihood values, introduced by \citet{sch1}. 

National Earthquake Hazard Reduction Program (NEHRP) gave a 
site classification scheme based on the 
average shear wave velocity for
upper 30 m soil column ($V_{S30}$). 
The standard engineering bedrock or the firm-rock site conditions
are considered to be more
realistic for the regional hazard computations \citep{nath2} and 
the standard engineering
bedrock corresponds to $V_{S30}$ $\sim$ $760$ m/s (defined as the boundary
site class BC). In the present study, 
the spatial distribution of probabilistic seismic hazard
in terms of Peak
Ground Acceleration (PGA) and $5\%$ damped Pseudo
Spectral Acceleration (PSA) at
different time periods for $10\%$ and $2\%$
probability of exceedance in $50$ years 
(corresponding to return periods $475$ and 
$2475$ years, respectively) at
engineering bedrock have been obtained.
The final hazard maps thus
produced are able to capture the spatial variations in seismic hazard
of Tripura. We expect that the results will be useful to structural 
engineers for earthquake resistant design of structures and helpful to 
government for taking decisions regarding disaster mitigation.

\section{Region of Study and Seismo-tectonic Framework}
\label{seistec}
\begin{figure}[h]
\begin{center}
	\rotatebox{0}{\includegraphics[scale=0.6]{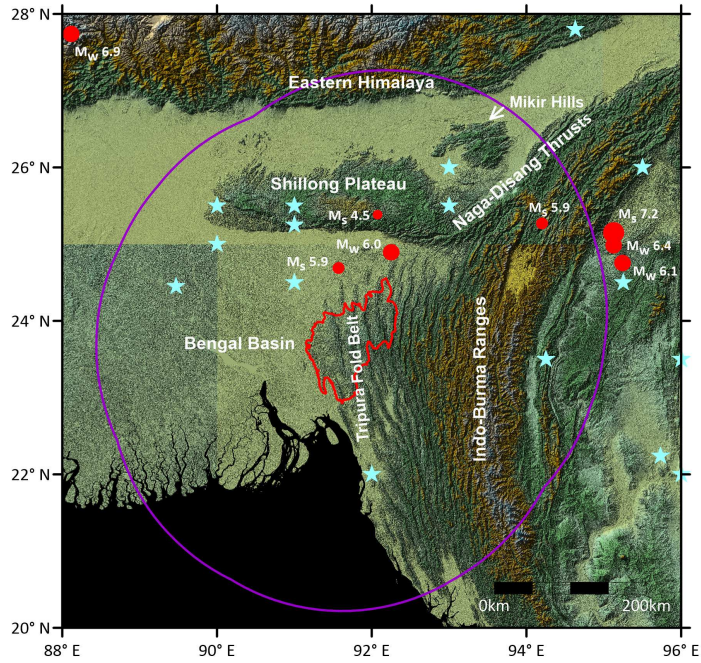}}
\caption{Region of the present study with the important geotectonic units. 
	The solid line in red colour shows the geographical boundary 
	(approximate) of Tripura and the solid line in
	purple colour shows a buffer of $300$ km. The events in asterix
	symbol indicate the earthquakes of magnitude $M_W > 7.0$. The
	events in filled circle are the strong-motion earthquakes considered
	for selection of suitable GMPEs.}
\label{ros}       
\end{center}
\end{figure}

The present study is focused on the seismic hazard assessment of Tripura,
one of the eight northeastern states in India. The region of study is
extended by $300$ km in radius from the geographical boundary of the state
of Tripura, $i.e$, a buffer of $300$ km is considered from 
the geographical boundary 
of Tripura. Therefore, the region of study ranges from $20\degree$ N - 
$28\degree$ N in latitude and $88\degree$ E - $96\degree$ E in longitude.
The region of study with the important geo-tectonic units 
is shown in Figure \ref{ros}. The hill shaded terrain representation 
SRTM data, used in Figure \ref{ros}, has been taken from Consortium 
for Spatial Information \citep{jar1}.

\begin{figure}[h]
\begin{center}
	\rotatebox{0}{\includegraphics[scale=0.6]{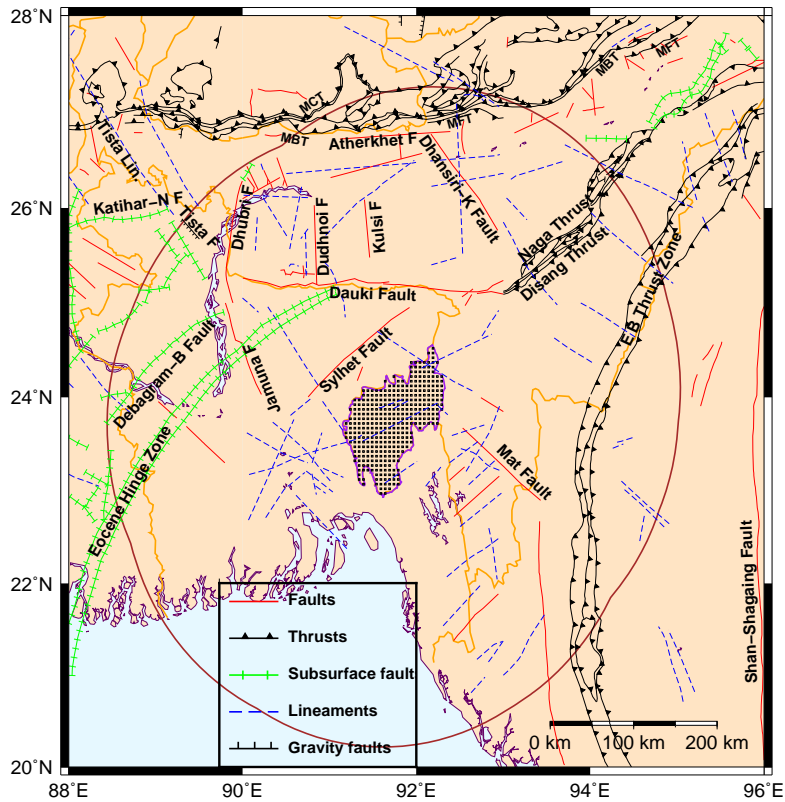}}
	\caption{Tectonic map of the study region. The international
        boundaries, coastlines and water bodies are produced from the
        Generic Mapping Tools (GMT) database.} 
\label{tecmap}       
\end{center}
\end{figure}

Northeast India is one of the most seismically active regions 
of the world \citep{kay1}. The seismotectonics of northeast India 
is summarized as south-directed over-thrusting from the north 
due to collision tectonics at the Himalayan Arc, and northwest 
directed over-thrusting from the southeast due to subduction 
tectonics at the Burmese Arc \citep{mukh1}. The study area covers 
Sheet numbers 13, 14, 16, 24, 25 and 26 of the Seismotectonic 
Atlas of India and Its Environs (SEISAT) published by 
Geological Survey India (GSI) \citep{dasg1}. It is 
represented by six major tectonic domains namely, the 
Himalayan mobile belt (Eastern Himalaya) and the Shillong 
Plateau-Mikir Hills in the north, the Naga-Disang Thrust 
system in the northeast, the Tertiary fold-thrust belt 
of the Indo-Burma Arc in the east, the outer molasse basin 
of Tripura-Chittagong in the south and  the Bengal Basin in 
the southeast.
The tectonic features (faults, thrusts, major lineaments etc.)
obtained from the SEISAT \citep{dasg1}, across the study region
is depicted in Figure \ref{tecmap}.

The Himalayan mobile belt is represented by four tectonic units
in the form of the crystalline 
complexes (F)F; the folded cover sequences (F)FC; the cover 
sequences (F)C of Buxa, Miri and Gondwana Groups; and the 
frontal belt (F) represented by Siwalik Group. 
These are separated by widespread thrust faults 
namely, Main Central Thrust (MCT), Main Boundary Thrust 
(MBT) and Main Frontal Thrust (MFT) and their subsidiary 
thrusts \citep{dasg1}. 
It is a pile of EW oriented thrust faults made of 
crystalline rocks of the great 
Himalaya.
These thrust faults have overrided 
the Indian Plate due to the northward underthrusting of Indian
Plate beneath southern Tibet \citep{gan1}. 
There is N-S convergence within the 
eastern Himalaya which is accommodated through thrust 
faulting on shallow north dipping fault plane within the 
eastern Himalayan wedge \citep{kum1}. The MCT 
separates two geologically distinct zones — the Lesser 
Himalaya to the south and the Higher Himalayan 
Crystallines to the north. The main discontinuity between 
the sub-Himalayan and lesser Himalaya zones is the MBT, 
presently marked by intense seismicity and disastrous 
earthquakes \citep{ang1}.

The Shillong Plateau consist mostly of Archaean Gneissic 
complex with Proterozoic intracratonic quartzite and 
phyllite of Shillong group intruded by acid and basic 
igneous rocks. During the Jurassic-Cretaceous period, 
the plateau rifted along the southern margin by the E-W 
steep north dipping Dauki Fault  and  earthquakes occur 
to depths of 30-50 km beneath the Shillong plateau 
\citep{bil1}. Shillong Plateau is characterized by a number 
of faults, shears and lineaments, which exhibit considerable 
seismicity. The great 1897 Shillong earthquake ($M_W$  8.0)
occurred in the Shillong Plateau which is noted as the 
largest magnitude earthquake observed in the study area.   
The north dipping Dauki Fault zone was considered responsible 
for the same \citep{old1}. 
The N-S striking Dudhnoi and Kulsi 
faults, the NE striking Barapani and Kalyani shear zones, 
and several NE-SW and N-S striking lineaments are the 
other major faults which contribute to the high seismic
activity in the Shillong Plateau. 
In the N-E of Shillong Plateau lies Mikkir hill which
is detached from Shillong plateau by seismically active 
NW-SE striking Kopili-North Dhansiri fault. 
The Shillong Plateau and Mikir Hills are identified as 
detached parts of the Indian Plate in the eastern 
Himalayan syntaxis \citep{gan1}. In the north, Shillong plateau 
is limited by Brahmaputra basin, which is characterized by 
several sets of neotectonic faults of which the NE-SW 
trending and E-W trending sets are the most conspicuous \citep{dasg1}.

The Naga-Disang thrust system forms a complex pattern with 
a narrow belt of imbricate thrust slices, known as the Belt 
of Schuppen. It is delineated from southeast of Brahmaputra 
foredeep by Naga thrust on the west and Haflong-Disang 
thrusts on the east. A number of NNW and NW trending lineaments 
cut across this over-thrust belt and traverse further south 
into the Paleogene inner fold belt \citep{dasg1}. The frontal thrust, 
although called the Naga thrust, is composed of many different 
thrusts. The uppermost thrust, known as Disang thrust, overrides 
all the lower thrusts in the north Cachar Hills. The Disang thrust 
passes westwards into Dauki fault.

The Indo-Burmese Arc  is a very important tectonic feature, 
and the important structural elements in this part of the 
region are the high angle reverse faults parallel to the 
regional north-south folds in the eastern part of the 
outer arc ridge within the Eastern Boundary Thrust Zone 
(EBTZ) and  the north westerly trending transverse 
Mat fault \citep{dasg1}. The Burmese Arc seismicity is an outcome 
of eastward under thrusting of the Indian Lithosphere below 
the Burma plate. \citet{sant1} illustrates that an inclined seismic 
zone, defining the Benioff zone, is present throughout Burma. 
The dip of the Benioff zone was found to $45\degree$ and the 
depth of penetration about 180km. Seismotectonic studies of 
the Indo-Burma region have been attempted by several 
investigators. An eastward dipping zone of seismicity, 
consistent with the Benioff zone in a typical subduction 
zone, has been inferred by \citep{guz1} who studied the focal 
mechanisms, focal depth distributions and geometry of 
Wadati-Benioff zone in this region and suggested that 
there are no interplate earthquake in this region.

The Tripura-Chittagong fold system is a typical 
fold-and-thrust belt with west-verging thrusts as 
a result of the eastward subduction of the Indian 
Plate \citep{ang1}. It constitutes a different tectonic domain 
of the Upper Tertiary Surma Basin belonging to the periphery 
of the Indo-Burma orogen.  It  comprises of Tripura-Mizoram 
Hills, The Cittagong hill tracts and  Coastal Burma, 
represents a large Neogene basin formed west of Paleogene 
Arakan-Yoma Fold Belt and is broadly confined within the Naga 
Thrust  and Dauki Fault to the north and Sylhet Fault 
and to the west and northwest. 
The Tripura-Chittagong fold system is composed of
narrow, long and doubly plunging folds. The N-S 
trending folds indicate eastward drag along
Dauki Fault in the north.  
The seismicity in 
this zone is sparse and the events are mostly shallow 
focused \citep{dasg1}.

The Bengal Basin is bordered on its west by the 
Precambrian basement complex of crystalline 
metamorphis of the Indian Sheid and to the 
east by the Tripura- Chittagong Fold Belt. 
The Bengal Basin comprises of enormous volume of 
sediments flown down by the Ganga-Brahmaputra 
drainage system building the world’s largest 
submarine fan, the Bengal fan. The Bengal fan 
conceals the northern extension of the Ninety 
Eastern Ridge and the bathymetric trench, 
related to subduction of Indian plate \citep{dasg1}.

Figure \ref{seistec} shows the seismicity map where the epicenters of the 
main shocks corresponding to different ranges of hypo-central depths
are superimposed on the tectonic map to correlate the past
seismicity with the identified tectonic features.
\begin{figure}[h]
\begin{center}
	\rotatebox{0}{\includegraphics[scale=0.6]{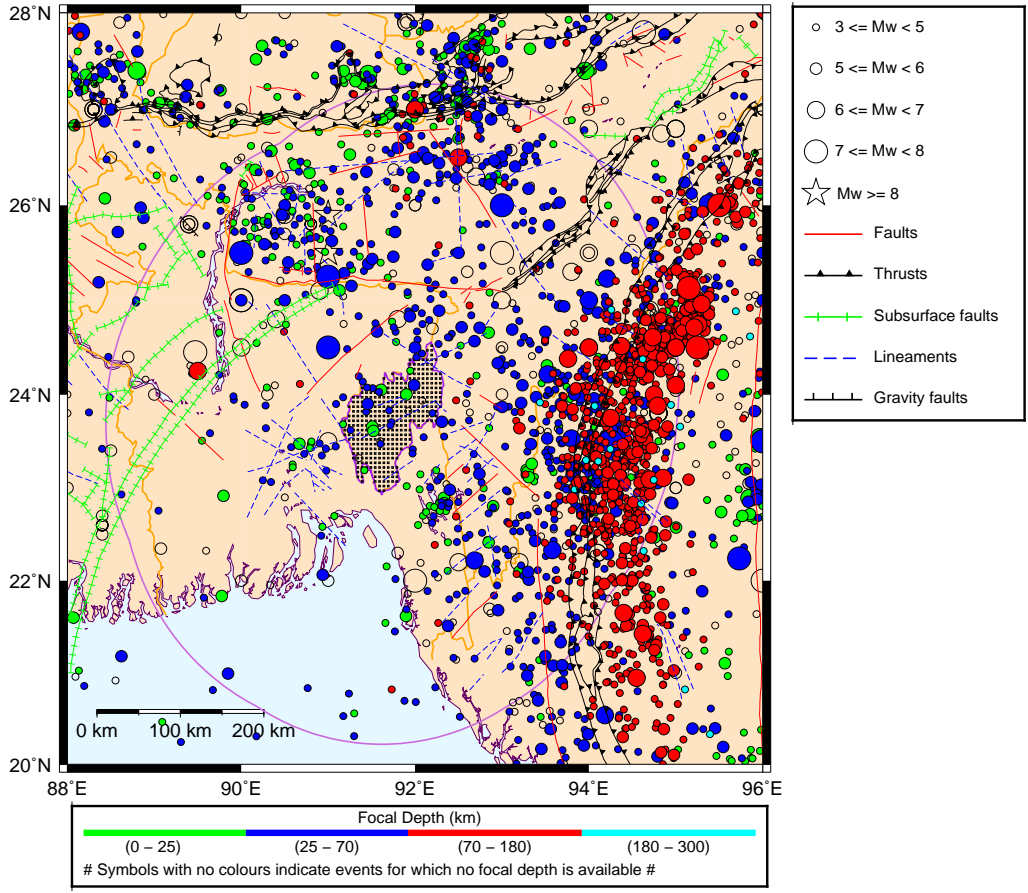}}
	\caption{Seismo-tectonic map of the study region depicting the
	epicenters of the main shocks on the tectonic features.} 
\label{seistec}       
\end{center}
\end{figure}

\section{Methodology and Computational Framework}
\label{mcf}
\subsection{Preparation of Earthquake Catalogue}
The region of the present study ranges from latitude $20 \degree$ N - 
$28 \degree$ N and longitude $88 \degree$ E - $96 \degree$ E. The first 
step of seismic hazard assessment is to prepare an earthquake catalogue.
Earthquake database which contains details of each earthquake event such
as time of occurrence in terms of year, month, day, hour and minute; 
location of occurrence in terms of local magnitude $M_L$, surface wave
magnitude $M_S$, body wave magnitude $m_b$ and moment magnitude $M_W$ is
known as earthquake catalogue. The earthquake catalogue for the present
analysis has been prepared from the reviewed  International Seismological 
Centre (ISC) bulletin, UK 
(\url{http://www.isc.ac.uk/iscbulletin/search/bulletin/}), 
National Earthquake Information Centre (NEIC), 
(\url{https://earthquake.usgs.gov/earthquakes/search/}), USGS and
Global Centroid Moment Tensor (GCMT) project supported by the National
Science Foundation (\url{https://www.globalcmt.org/CMTsearch.html})
for instrumental period and for pre-instrumental and early instrumental 
period, the database has been taken from various published sources 
\citep{jai1,amb1,old2},
supplemented with data from India Meteorological Department (IMD), New Delhi.
The catalogue compiled for the present study contains a total of $4491$
events starting from 825 A.D. The compiled catalogue contains magnitude in 
different scales ($M_L$, $M_S$, $m_b$ and $M_W$). Conversion of different 
magnitude scales into one type of magnitude ($M_W$) using suitable conversion
relations is required to be done for seismic hazard analysis because most of 
the GMPEs are developed in terms of $M_W$ to avoid saturation effects. The 
process of conversion is termed as homogenization. The homogenization has been
carried out using global empirical relations \citep{fliz1,scor1,sipk1}. 
The preference or the
priority of the type of magnitude taken for homogenization is $M_W$ 
$\rightarrow$ $M_S$ $\rightarrow$ $m_b$ $\rightarrow$ $M_L$. Various 
conversion relations used in the present study is given in Table 
\ref{tab:conrel}.

\begin{table}[h]
\caption{Conversion relations used in the present study}
\label{tab:conrel}       
\begin{tabular}{cccc}
\hline\noalign{\smallskip}
Type of \\Magnitude & Magnitude range(s) & Conversion relation(s) & Reference(s)\\
\noalign{\smallskip}\hline\noalign{\smallskip}
 & $3.0$ $\leq$ $M_S$ $<$ $6.2$ & $M_W=0.67M_S+2.07$ & \citet{scor1}\\
$M_S$ & $6.2$ $\leq$ $M_S$ $\leq$ $8.2$ & $M_W=0.99M_S+0.08$ & \citet{scor1}\\
& $M_S$ $>$ $8.2$ & $M_W=0.8126M_S+1.1723$ & \citep{fliz1}\\
\noalign{\smallskip}\hline\noalign{\smallskip}
& $3.5$ $\leq$ $m_b$ $\leq$ 5.5 & $M_W=0.85m_b+1.03$ & \citet{scor1}\\
$m_b$ & $5.5$ $<$ $m_b$ $\leq$ $7.3$ & $M_W=1.46m_b-2.42$ & \citet{sipk1}\\
& $m_b$ $>$ $7.3$ & $M_W=1.0319m_b+0.0223$ & \citep{fliz1}\\
\noalign{\smallskip}\hline\noalign{\smallskip}
& $M_L$ $\leq$ $6.0$ & $M_W=M_L$ & \citet{heat1}\\
$M_L$ & $M_L$ $>$ $6.0$ & $M_W=0.08095M_L+1.30003$ & \citep{fliz1}\\
\noalign{\smallskip}\hline
\end{tabular}
\end{table}

The seismic events in the catalogue contain main shocks and triggered events
(foreshocks and aftershocks). Main shocks are statistically independent and
follow Poissonian distribution. Triggered events are dependent on main shocks
and tend to cluster in space and time. Seismic hazard analysis is generally
based on the assumptions of Poissonian distribution of earthquakes. Therefore
it is necessary to identify and remove the foreshocks and the aftershocks 
from the catalogue. The process to eliminate the dependent events from the
earthquake catalogue is known as declustering. The \citet{Gard1} 
window method for
identification of dependent events is widely used in practical seismic 
hazard analysis and the same is adopted here. This approach states that 
aftershocks are dependent (a non-Poissonian process) on the size of the
main shocks and the dependent events need to be removed in accordance with 
defined distance and time windows. The time window $T(M)$ in days and 
distance window $L(M)$ in km after \citet{Gard1} are as follows:
\begin{equation*}
	L(M)=10^{0.3238M+0.983}
\end{equation*}
\begin{equation}
	\label{eq:gk}
	T(M)=
	\begin{cases}
		10^{0.032M+2.7389} & \text{if $M$ $\geq$ $6.5$} \\
		10^{0.032M+2.7389} & \text{if $M$ $<$ $6.5$}
	\end{cases}
\end{equation}
For any earthquake of magnitude $M$ in the catalogue, the subsequent shocks 
are identified as aftershock if they occur within the time window $T(M)$ and
the distance window $L(M)$. It is more practical to put an upper limit on the
magnitude of aftershocks, say at least one unit magnitude below the magnitude
of the main shock \citep{bath1}. The method can also be applied to identify the
foreshocks, if the time of occurrence is before the time of the shock under
consideration and it has not been identified as aftershock of an earlier 
main shock. No upper limit is required to be put on the magnitude of the
foreshock, except that it should be lower than that of the main shock. The
declustered earthquake catalogue contains a total of $2694$ main shocks in 
$M_W$ unit, $i. e$, declustering eliminates about $40$$\%$ of the events in
the catalogue.

\subsection{Delineation of Seismogenic Source Zones}
\begin{figure}[!h]
\begin{center}
\begin{tabular}{cc}
      \resizebox{55mm}{!}{\rotatebox{0}{\includegraphics[scale=0.6]{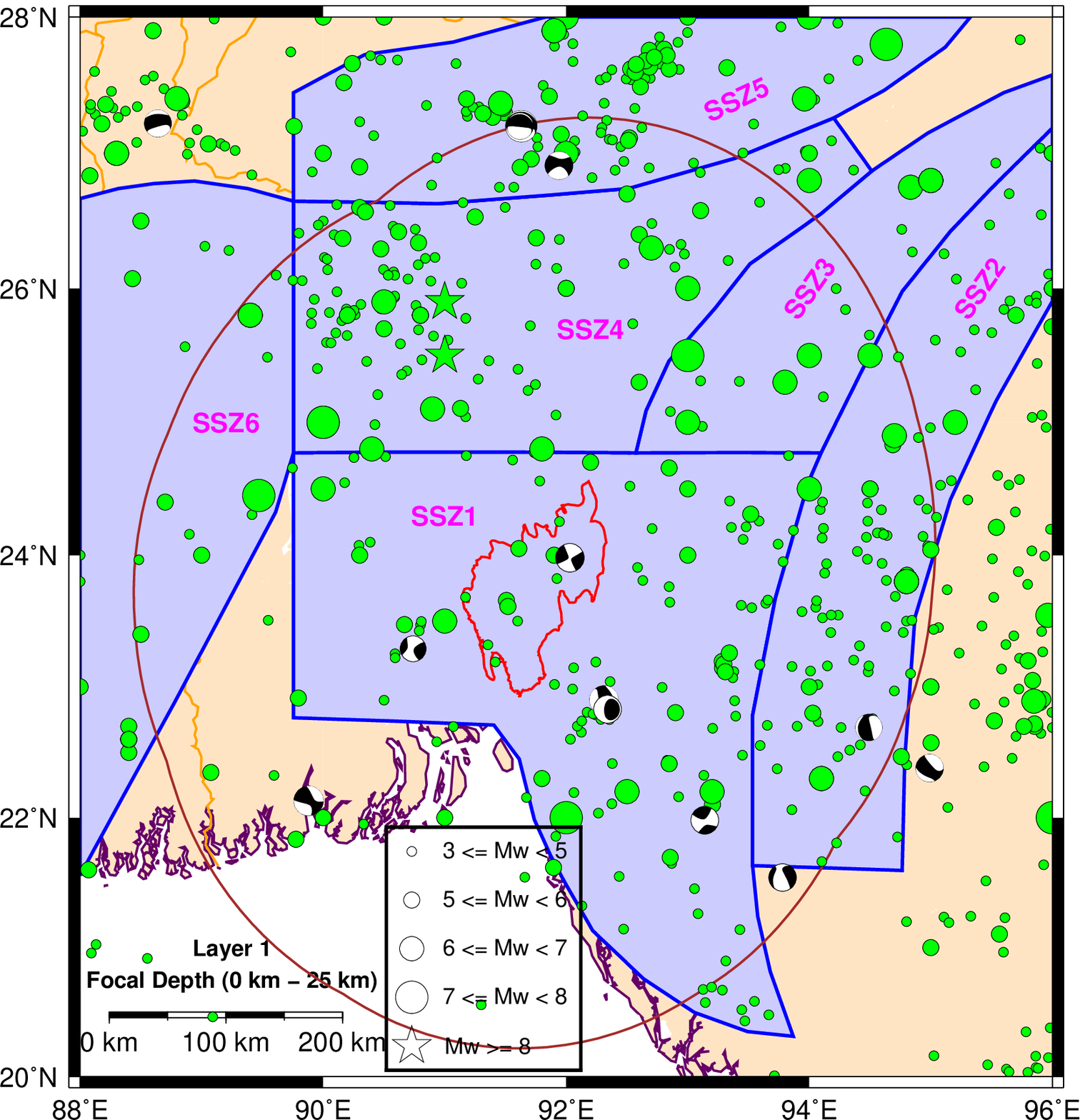}}} &
      \resizebox{55mm}{!}{\rotatebox{0}{\includegraphics[scale=0.6]{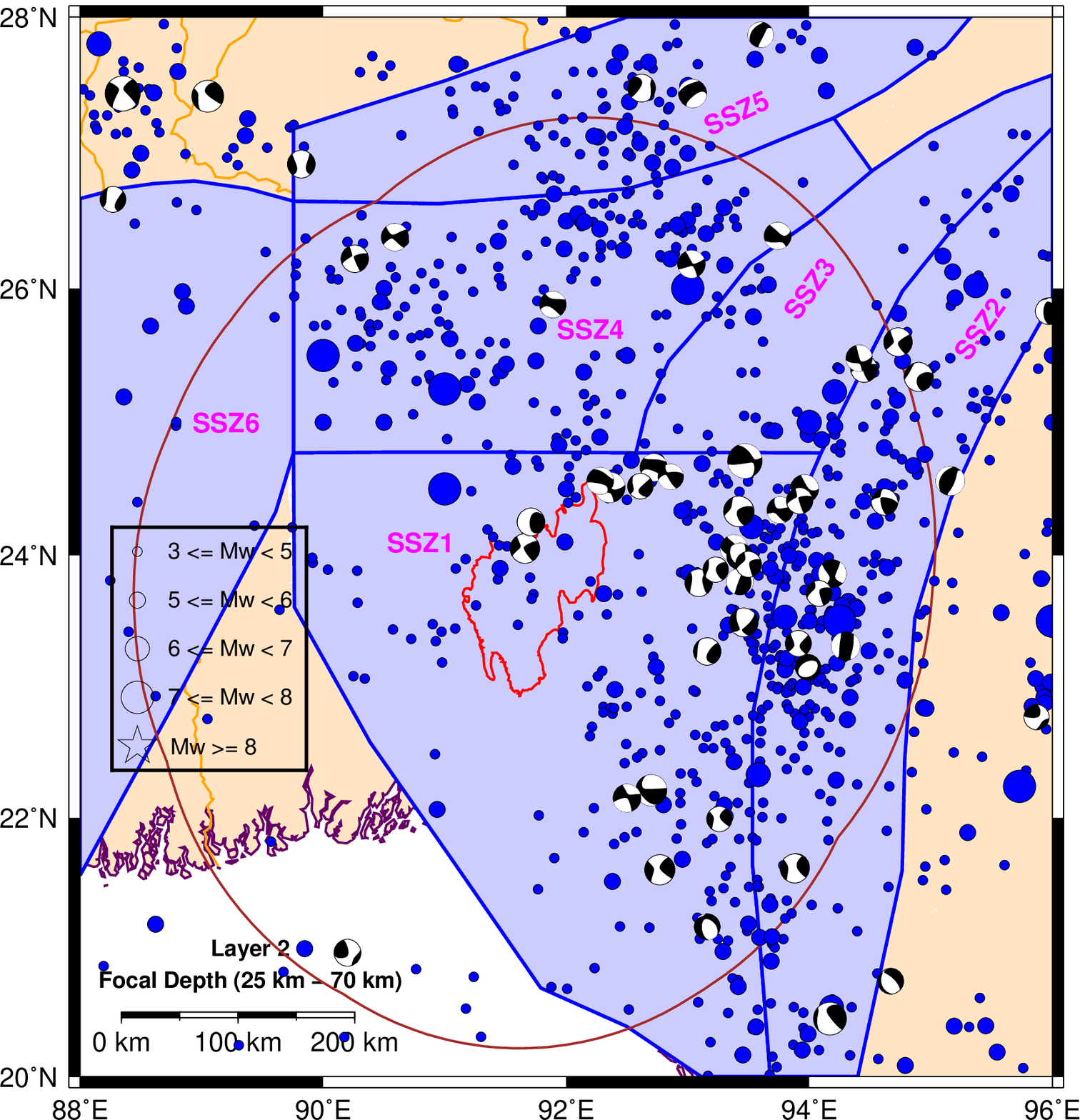}}} \\
      \resizebox{55mm}{!}{\rotatebox{0}{\includegraphics[scale=0.6]{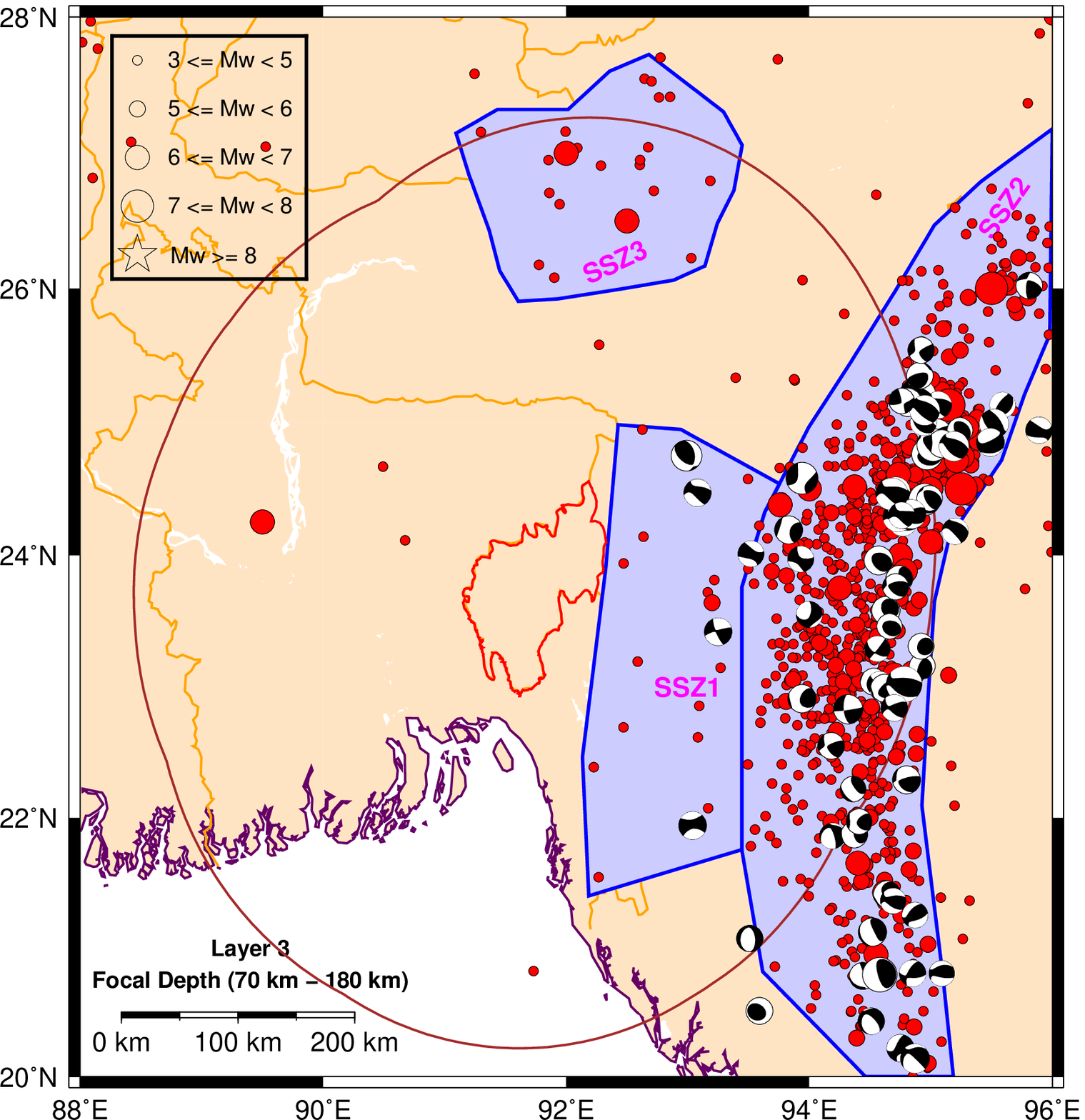}}} &
      \resizebox{68mm}{!}{\rotatebox{0}{\includegraphics[scale=0.6]{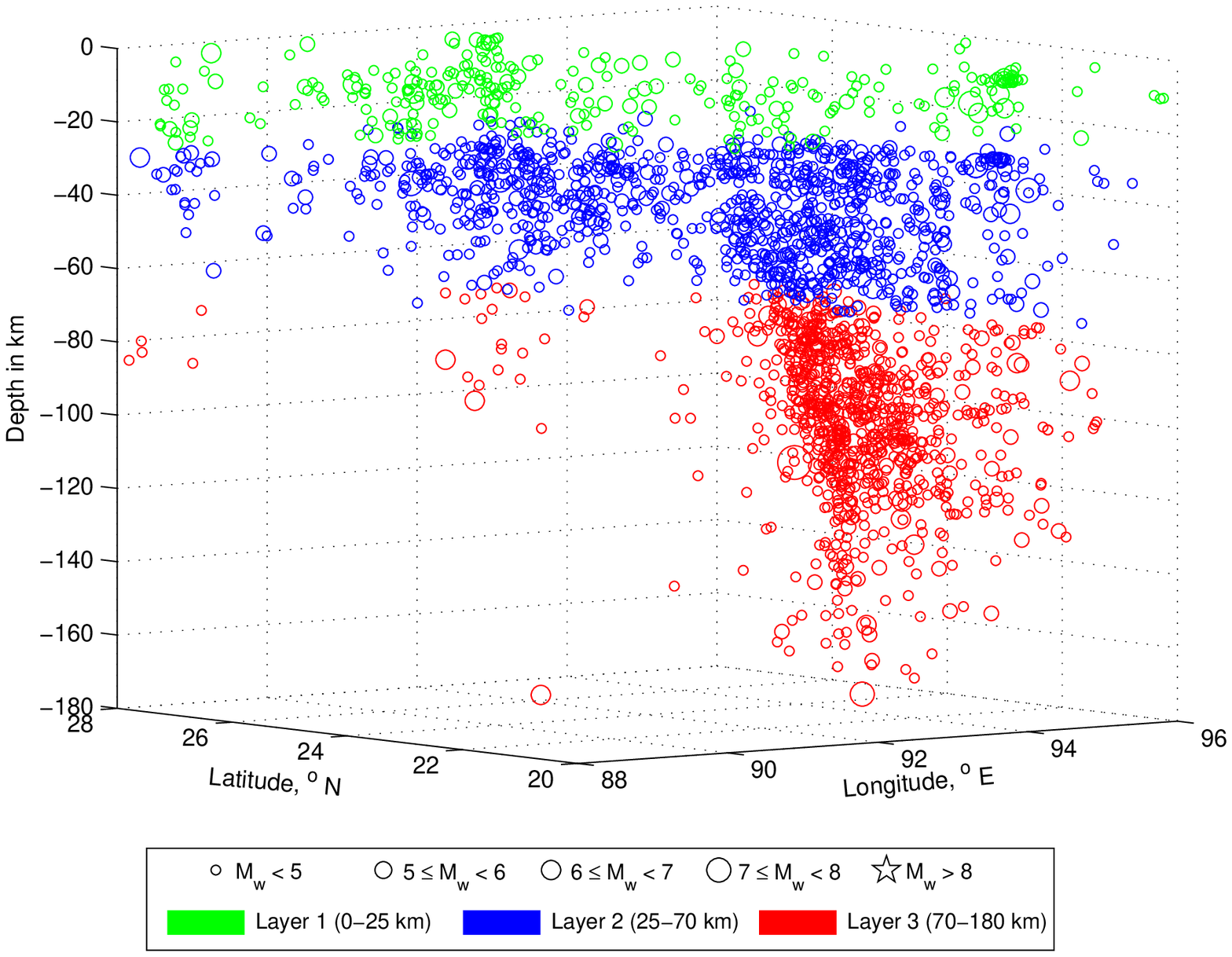}}} \\ 
    \end{tabular}
\end{center}
\caption{Seismogenic source zones in the study area for all the three Layers
	along with the 3D depth-section of the main shocks.}
\label{ssz}
\end{figure}

Identification of seismogenic sources is one of the most
important steps in seismic hazard assessment. Due to plate
boundary activity in northern Himalaya, intraplate activity in
Shillong Plateau and subduction process in Indo-Burmese range,
the seismotectonic setup observed in the study region is very
complex as described in Section \ref{intro}. Source delineation is 
primarily based on tectonic trends and seismicity of the region.. However,
there are not any consistent criteria for defining seismogenic source 
zone (SSZ) so far. We found the methodology given by \citep{nath2,nath4} to be
appropriate and has been adopted. They suggested layered polygonal
SSZs corresponding to different ranges of hypo-central depths. It has 
been observed by various researchers that seismicity patterns and
source dynamics have significant variation with depths 
\citep{alle1,chri1,tsap1}. 
Therefore the consideration of single set of seismicity parameters over
the entire depth ranges may result in inaccurate estimation of seismic 
hazards. Based on the hypo-central depth distribution of seismicity in the
study region, three hypo-central depth ranges (in km) corresponding to
$0-25$ (Layer 1), $25-70$ (Layer 2) and $70-180$ (Layer 3) respectively
are considered and SSZs are delineated on the basis of seismicity patterns, 
fault networks and similarity in the style of focal mechanisms for all the
three Layers as depicted in Figure \ref{ssz}. The epicenters of the main 
shocks from the declustered catalogue and the focal mechanisms for earthquakes
of magnitude $\geq 5.0$ $M_W$, extracted from the GCMT database, for respective
layers are also shown in the Figure \ref{ssz}. The three dimensional (3D)
depth-section of the 
main shocks is plotted in the bottom right panel of Figure \ref{ssz}.
Layer $3$ mostly corresponds to subduction process. The SSZs are described 
briefly in the following.

A SSZ is considered as an area of diffused seismicity with 
distinctly different seismogenic potential in terms of the maximum magnitude as 
well as the occurrence rate of earthquakes in different magnitude ranges.  
Considering the highly complex spatial distribution and correlation of seismic 
activity with the tectonic features in the study area, six broad SSZs have been 
identified for each hypo-central Layer $1$ and $2$ and three SSZs for hypo-central 
Layer $3$ as shown in Figure \ref{ssz}.

\noindent\underline{SSZs for Hypo-central Layers 1 and 2}

The geometrical coverage of SSZs for
Layer $1$ and $2$ are similar.
The SSZ$1$ is occupied by frontal Indo-Burma fold belt in the east having lower 
level of seismicity. The northern part of Bengal Basin is covered in the west. 
The central part, occupied by N-S trending Tripura-Cachar Fold Belt, constitutes 
a domain of the upper territory Surma Basin belonging to the periphery of the 
Indo-Burma orogen. The state boundary of Tripura lies in this zone. 
The SSZ$2$ encompasses intense seismicity related to the Territory fold thrust 
belt of the Indo-Burma Arc caused by the subduction of the Indian plate below 
the Burmese plate.  
The SSZ$3$ lies southeast of Brahmaputra Basin. The foredeep exposes a narrow 
belt of imbricate thrust slices, the Shuppen Belt. This belt is delineated on 
the west by Naga Thrust and on the east by Haflong-Disang Thrust. 
The SSZ$4$  is  seismically active which encompasses mainly Shilling Plateau 
and part of the Brahmaputra valley in the northeast. The northern margin of 
the Sillong Plateau subsided step wise to form the basement for the deposition 
of upper territory sediments and huge pile of alluvium in the Brahmaputra 
foredeep extending up to Himalayan frontal belt. To the south and east the 
plateau terminates along Haflong-Disang Thrust. 
The SSZ$5$  forms the easternmost segment of Himalaya, consists of major thrusts 
like Main Central Thrust (MCT), Main Boundary Thrust (MBT), 
and Main Frontal Thrust (MFT), along with several secondary thrusts and transverse 
tectonic features. 
The SSZ$6$  consists of the Indo Gangetic plains including Himalayan 
foredeep and shows somewhat subdued seismic activity. 

\noindent\underline{SSZs for Hypo-central Layer $3$}

The SSZ$1$ encompasses low magnitude deep events in the  
frontal Indo-Burma fold belt. 
The SSZ$2$ includes intense seismicity related to the Tertiary 
fold - thrust belt of the Indi-Burma Arc caused by the 
subduction of the Indian plate 
below the Burmese plate and the events have deeper focal depths. 
The SSZ$3$ includes low magnitude deep events from  northern part 
of Shillong Plateau and Eastermost segment of Himalaya consists of 
MCT, MBT and MFT. 

\subsection{Catalogue Completeness}
\begin{figure}[h]
\begin{center}
	\rotatebox{0}{\includegraphics[scale=0.8]{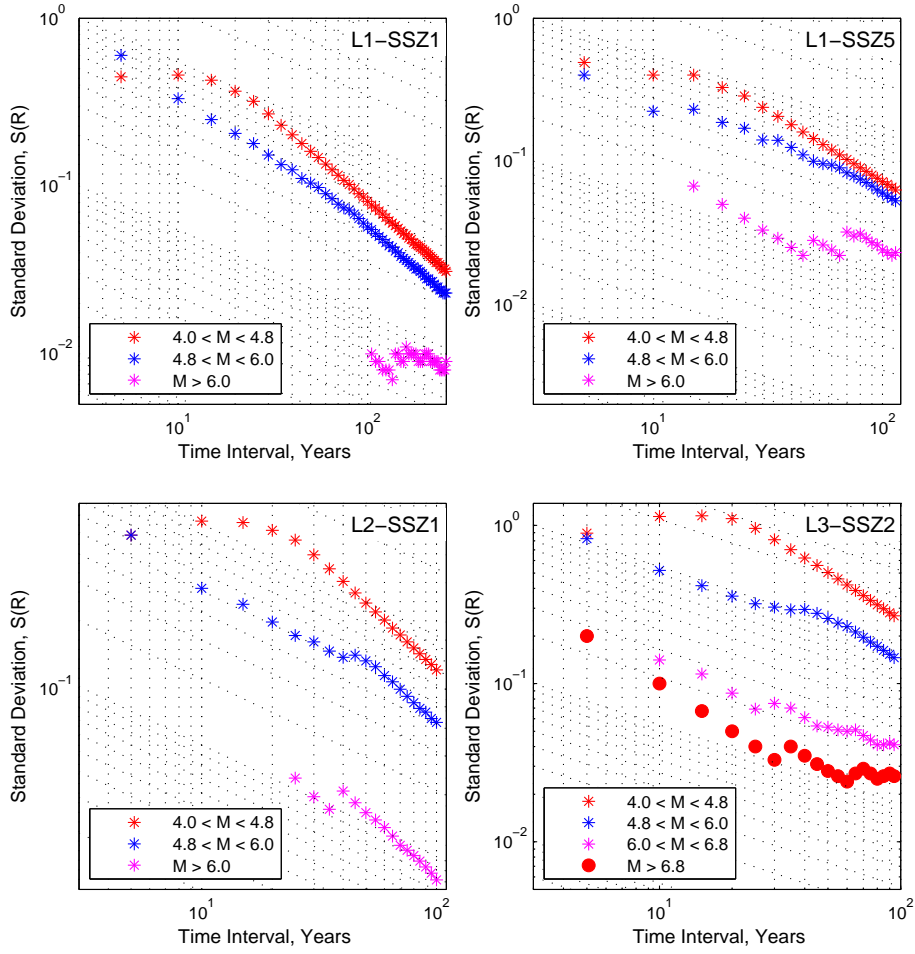}}
\caption{Typical completeness plots for SSZs $1$ and $5$ in Layer $1$,
	SSZ $1$ in Layer $2$ and SSZ $2$ in Layer $3$.}
\label{stepp}       
\end{center}
\end{figure}

It is recognized that earthquake data in the catalogue are generally incomplete for 
smaller magnitude earthquakes in the early time due to inadequate instrumentation.
However, due to short return periods of smaller magnitude, their occurrence rates
can be evaluated even from recent data, say, approximately for last $25$ years. However,
to have a reliable estimate for the occurrence rates of larger magnitude earthquakes
with long return period, the data for a much longer period need to be considered.
Failure to correct for data incompleteness may result in underestimation of the mean
rates of occurrence of earthquake. The correction can be done by identifying the time
period of complete data for a pre-defined magnitude ranges. Reliable mean rates of
occurrence of earthquake for the said magnitude ranges can then be computed from the 
complete data. \citet{step} proposed a statistical method to calculate the period of 
completeness and this method has been followed in the present analysis to determine 
the period of completeness. Stepp suggested that if the earthquakes in a catalogue 
are reported completely, they will follow a Poissonian distribution with constant 
occurrence rates. If $R(M)$ is the average number of events per year for a specific
magnitude range centered about $M$ for a time interval of $T$ years, the standard
deviation $S_R$ of of $R(M)$ is given by
\begin{equation}
	S_R=\sqrt{R(M)/T}
	\label{catcom}
\end{equation}
The stationarity of $R(M)$ guarantees that $S(R)$ behaves as $1/\sqrt{T}$. From the
plot of $S(R)$ versus $1/\sqrt{T}$, known as 'completeness plot', the period of
completeness is determined by a marked deviation of of the $S(R)$ values from the 
linearity of the $1/\sqrt{T}$ slope. The period of completeness becomes successively 
longer with higher magnitude range. Typical completeness plots for SSZs $1$ and $5$ in 
Layer $1$, SSZ $1$ in Layer $2$ and SSZ $2$ in Layer $3$ are shown in 
Figure \ref{stepp}.

\subsection{Maximum Earthquake Prognosis}
The maximum earthquake ($M_{\tt max}$) is the largest seismic event 
characteristic of the terrain under seismotectonic and stratigraphic 
conditions. There are a number of methods for determining $M_{\tt max}$
values \citep{well1,kijk1,kijk2} and a certain degree of subjectivity is always 
associated with each method \citep{boll1}. For example, \citet{well1} 
method needs
the fault rupture length of the future earthquakes to be specified and
this cannot be determined with any confidence due to lack of scientific
basis. In the present study, a maximum likelihood method for maximum
earthquake estimation, known as Kijko-Sellevoll-Bayesian (KSB) method,
has been adopted for its wide acceptability and sound mathematical
background behind it. From the knowledge of the Gutenberg-Richter (GR)
probability distribution function (PDF), it is possible to construct 
the Bayesian version of the Kijko-Sellevoll (KS) estimator of 
$M_{\tt max}$ \citep{kijk3} and the KSB estimator of $M_{\tt max}$ can
be calculated by iterative process. 
Maximum earthquake 
prognosis has been performed for all the SSZs of all the three layers.

\subsection{Estimation of Seismicity Parameters}
\begin{figure}[h]
\begin{center}
	\rotatebox{0}{\includegraphics[scale=0.6]{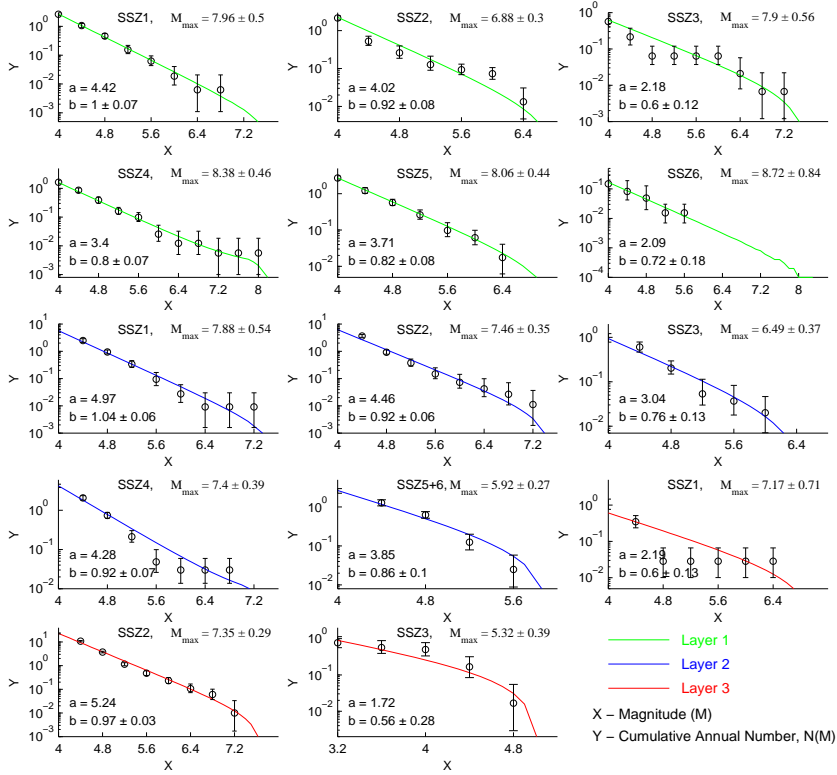}}
\caption{Magnitude-frequency distribution plots for main shocks in each of the
	polygonal SSZs in all the three Layers.}
\label{recur}       
\end{center}
\end{figure}

The evaluation of seismicity parameters is considered to be the most important
for hazard estimation. Earthquake occurrences across the the globe are
accepted to follow an exponential distribution, given by \citet{guten1} as
\begin{equation}
	\log N(M)=a-bM
	\label{gr}
\end{equation}
where $N(M)$ is the mean annual rate of exceedance for magnitude $M$,
$i.e$, cumulative number of events with magnitude equal to or 
greater than $M$. The value of $10^a$ is the mean yearly number of earthquakes of
magnitude equal to or greater than zero and $b$ (or the $b$ value)
describes the relative likelihood of large and small earthquakes. The
$a$ and $b$ values are generally obtained by regression analysis of data
available for a SSZ of interest. The standard Gutenberg-Richter (GR) 
relation covers an infinite range of magnitude ranging from 
$-\infty$ to $+\infty$. For engineering purposes, the effects of very
small earthquakes are of little interest and it is common to disregard
those that are not capable of causing significant damage \citep{kra1}.
It is therefore necessary to put a lower bound, $i.e$, to consider a 
lower threshold ($M_{\tt min}$) on the magnitude. At the other end of
the magnitude scale, the standard GR law predicts nonzero mean rates of
exceedance for magnitude up to infinity. However, there is always an 
upper bound magnitude or a maximum magnitude ($M_{\tt max}$) 
associated with all the SSZs. With the imposition of a lower bound and
an upper bound magnitude, the GR law is modified to a truncated 
exponential distribution in the magnitude range $M_{\tt min}$ to
$M_{\tt min}$ and is given by \citep{corn1}
\begin{equation}
	N(M)=N(M_{\tt min})\frac{e^{-\beta(M-M_{\tt min})}-e^{-\beta(M_{\tt max}-M_{\tt min})}}
	{1-e^{-\beta(M_{\tt max}-M_{\tt min})}}, ~M_{\tt min} < M < M_{\tt max}
	\label{modgr}
\end{equation}
where $\beta=b\ln 10$. Although the choice of $M_{\tt min}$ is not crucial in
Eqn.~\ref{modgr}, a suitable $M_{\tt min}$ is necessary for computation of
hazard \citep{bomm1}. 
$M_{\tt min}$
is taken as $4.0$ in the present study. 
The GR relationship for a SSZ is fitted using the maximum likelihood method 
of \citet{weic1} by first identifying the periods of completeness for 
different magnitude ranges.
As the number of events in SSZ $6$ of 
Layer $2$ is very less, SSZ $5$ and $6$ of Layer $2$ are merged together to get 
a good fit by the bounded recurrence law.
The bounded recurrence law of Eqn.~\ref{modgr}
cannot represent the magnitude-frequency dependence of SSZ $4$ of both 
Layer $1$ and Layer $2$ well. 
Therefore a more complex recurrence law, known as characteristic earthquake
recurrence law developed by \citet{youn2} is applied for those two SSZs. The
characteristic earthquake model predicts higher rates of exceedance at magnitudes
near the characteristic earthquake magnitude and lower rates at lower magnitudes.
The seismicity parameters estimated by the applicable recurrence relationship for all
the polygonal SSZs in different Layers are listed in Table \ref{seispar} and the 
magnitude-frequency distribution plots in each of the SSZs in different Layers are
shown in Figure \ref{recur}. 
\begin{table}[h]
\begin{center}
\caption{Estimated seismicity parameters for all the polygonal SSZs}
\label{seispar}       
\begin{tabular}{cccccc}
\hline\noalign{\smallskip}
 & SSZ(s) & $b$-value(s) & $a$-value(s) & $M_{\tt max}$ & $M_{\tt max}^{\tt obs}$ \\
\noalign{\smallskip}\hline\noalign{\smallskip}
 &1  & 1.00 $\pm$ 0.07  & 4.42  & 7.96 $\pm$ 0.50 & 7.53  \\
 &2  & 0.92 $\pm$ 0.08  & 4.02  & 6.88 $\pm$ 0.30 & 6.71 \\
Layer 1 &3  & 0.60 $\pm$ 0.12  & 2.18 & 7.90 $\pm$ 0.56 & 7.40  \\
 &4  & 0.80 $\pm$ 0.07  & 3.40 & 8.38 $\pm$ 0.46 & 8.00  \\
 &5  & 0.82 $\pm$ 0.08 & 3.71 & 8.06 $\pm$ 0.44 & 6.72 \\
 &6  & 0.72 $\pm$ 0.18 & 2.09 & 8.72 $\pm$ 0.84 & 7.92 \\
\noalign{\smallskip}\hline\noalign{\smallskip}
	&1 &1.04 $\pm$ 0.06  &4.97 &7.88 $\pm$ 0.54 &7.40  \\
 &2 &0.92 $\pm$ 0.06 & 4.46 & 7.46 $\pm$ 0.35 & 7.21  \\
Layer 2 &3 & 0.76 $\pm$ 0.13 & 3.04 & 6.49 $\pm$ 0.37 & 6.22  \\
 &4 & 0.92 $\pm$ 0.07 & 4.28 & 7.40 $\pm$ 0.39 & 7.10  \\
 &5+6 & 0.86 $\pm$ 0.01 & 3.85 & 5.92 $\pm$ 0.27 & 5.82 \\
\noalign{\smallskip}\hline\noalign{\smallskip}
 &1 & 0.60 $\pm$ 0.13 & 2.19 & 7.17 $\pm$ 0.71 & 6.51  \\
Layer 3 &2 & 0.97 $\pm$ 0.03 & 5.24 & 7.35 $\pm$ 0.29 & 7.20  \\
 &3 & 0.56 $\pm$ 0.28 & 1.72 & 5.32 $\pm$ 0.39 & 5.02 \\
\noalign{\smallskip}\hline
\end{tabular}
\end{center}
\end{table}

\subsection{Smooth-gridded Seismicity}
The seismogenic source is formulated with two schemes, namely smooth-gridded seismicity
and uniform-seismicity areal zones (or uniformly smoothed). The former entails spatially
varying annual activity rates while $b$-value and $M_{\tt max}$ remain fixed within the
source zones. This assumes $b$-value and activity rate to be uncorrelated and a 
non-uniform distribution of earthquake probability within a zone 
\citep{nath2}. On the other
hand, the uniform areal seismicity postulates each point within the zone to have equal
probability of earthquake occurrences.

\begin{figure}[!h]
\begin{center}
\begin{tabular}{cc}
      \resizebox{60mm}{!}{\rotatebox{0}{\includegraphics[scale=0.6]{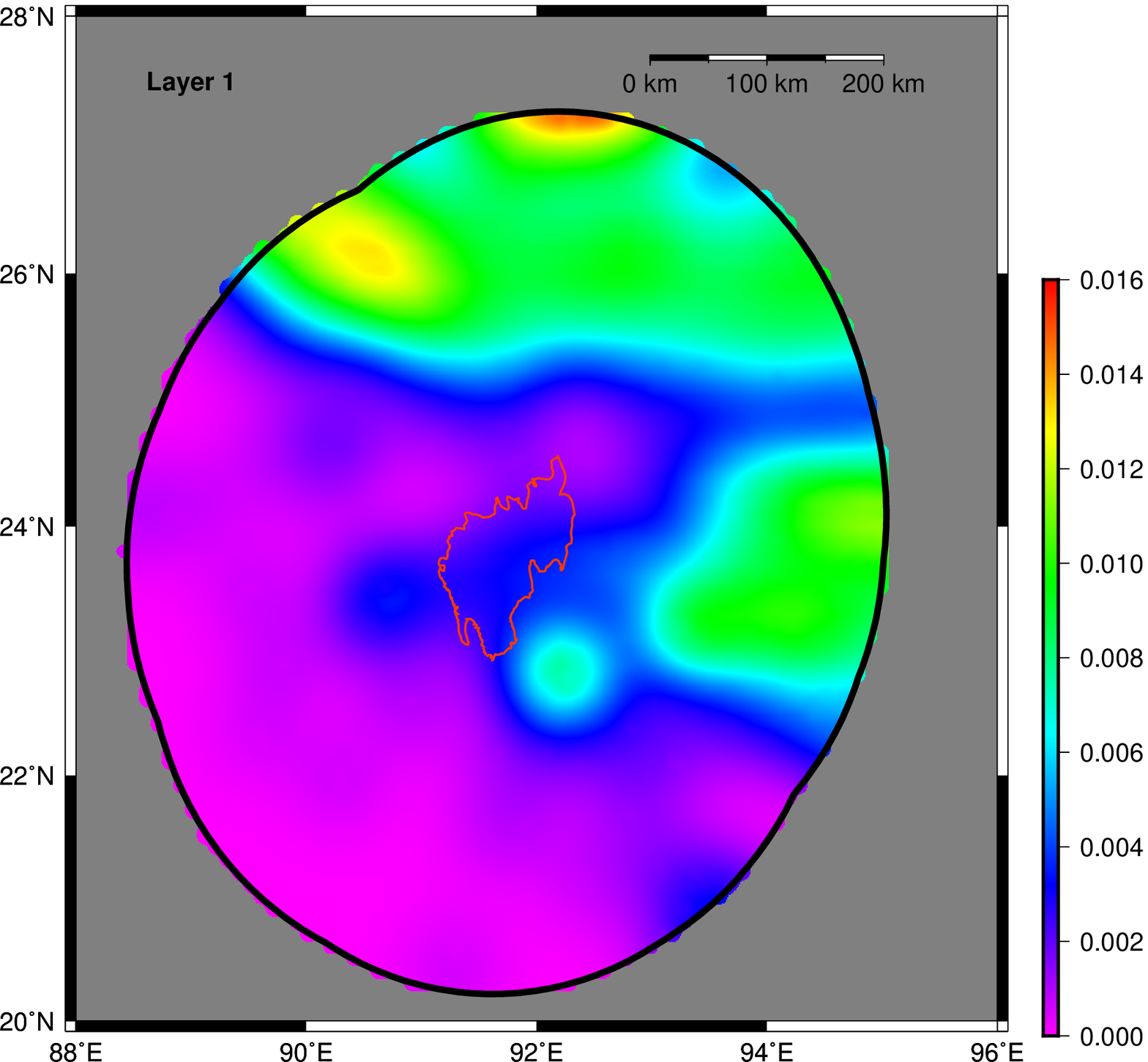}}} &
      \resizebox{60mm}{!}{\rotatebox{0}{\includegraphics[scale=0.6]{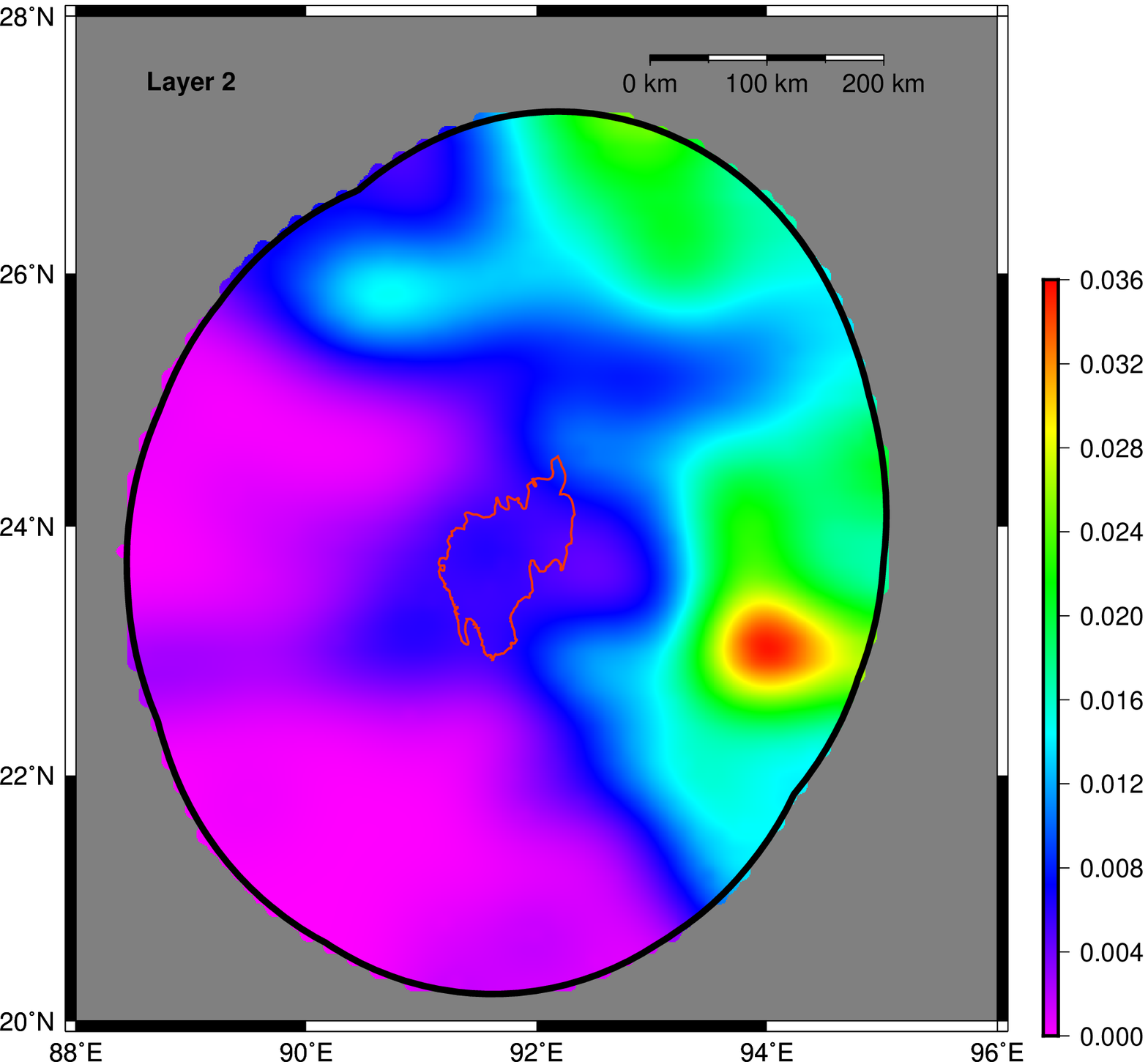}}} \\
      \resizebox{60mm}{!}{\rotatebox{0}{\includegraphics[scale=0.6]{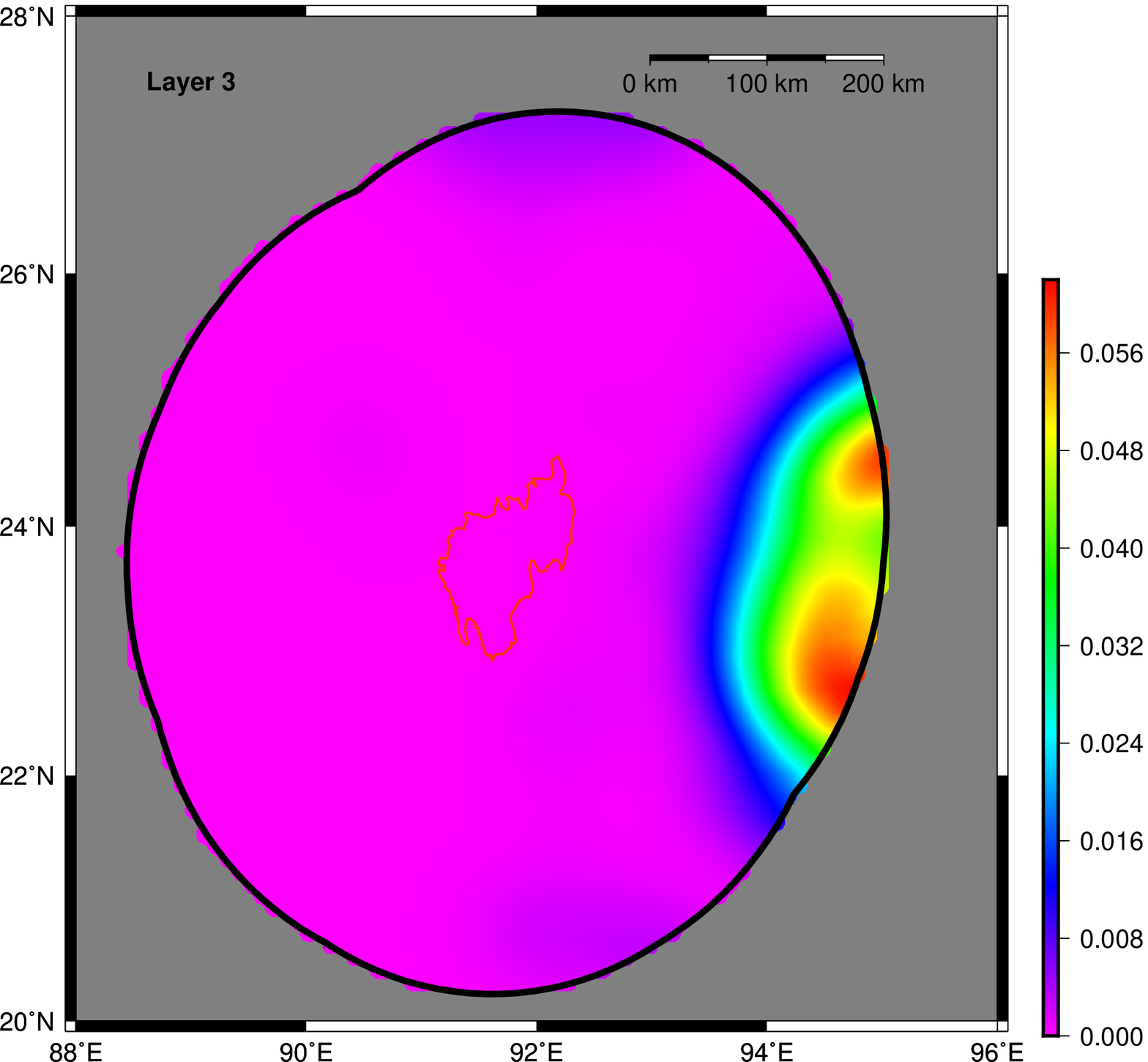}}} &
    \end{tabular}
\end{center}
\caption{Smooth-gridded seismicity analysis for threshold magnitude $4.0$ $M_W$
	for all the three Layers.}
\label{sgs}
\end{figure}

The contribution of background events for hazard perspective is calculated using 
smooth-gridded seismicity model. It allows modeling of discrete earthquake 
distributions into spatially continuous probability distributions. The technique
given by \citet{fran1} is employed for this purpose in the present study. The
technique has been previously employed by several researchers \citep{fran2,stir1,
lap1,jai2,nath4,nath2}. In the present analysis, the study region is gridded at a
regular interval of $0.05\degree$$\times$$0.05\degree$. The smoothened function
is given as follows:
\begin{equation}
	N_i(m_r)=\frac{\sum\limits_{j} n_j(m_r)e^{-(d_{ij}/c})^2}
	{\sum\limits_{j} e^{-(d_{ij}/c})^2}
\label{sgseq}
\end{equation}
where $n_j(m_r)$ is the number of events with magnitude $\geq m_r$,
$d_{ij}$ is the distance between $i^{th}$ and $j^{th}$ cells and 
$c$ denotes the correlation distance which characterizes uncertainty in the
epicentral location and is assumed to be $50$ km in the present analysis. 
The sum is calculated in cells $j$ within a distance of $3c$ of cell $i$. The
annual activity rate $\lambda_{m_r}$ is computed as $N_i(m_r)/T$ where $T$ is
the (sub)catalogue period for threshold magnitude $4.0$ $M_W$. The subcatalogue
for the threshold magnitude $4.0$ $M_W$ covers the period $1988 - 2020$. The
smooth-gridded seismicity for threshold magnitude $4.0$ $M_W$ for all the Layers
are depicted in Figure \ref{sgs}. From the smooth-gridded seismicity analysis, the
areas of probable asperities can be identified \citep{nath4}.

\subsection{Selection of GMPEs}
Appropriate selection and ranking of GMPEs against recorded 
strong motion data are critical for a successful logic tree 
implementation in the PSHA for incorporating the epistemic
uncertainties. A quantitative evaluation of the performance 
of the selected GMPEs against the observation (the recorded
strong motion data) is required for this purpose. To quantify
the level of agreement between the observations and the
predictions, a number of statistical measures of the 
goodness-of-fit of an equation to a set of data are calculated
in the present analysis following the approach proposed by
\citet{sch1}. This quantitative suitability is often referred to
as ``efficacy test" of a GMPE for a particular region. These
goodness-of-fit measures include mean, median and standard
deviation of the normalized residual as well as the median
values of the likelihood parameter after \citet{sch1}. The
normalized residual is given by
\begin{equation}
	Z=\frac{\log(\text {GM})_{\tt obs}-\log(\text {GM})_{\tt predicted}}
	{\sigma_{\tt T}}
\label{nr}
\end{equation}
where $(\text {GM})_{\tt obs}$ and $(\text {GM})_{\tt predicted}$ 
are the observed and the predicted ground motions and
$\sigma_{\tt T}$ is the total standard deviation of the GMPE
used. So normalized residual ($Z$) represents the distance of the
data from the logarithmic mean measured in units of
$\sigma_{\tt T}$. The likelihood parameter is calculated as
\begin{equation}
LH(\vert Z \vert)= Erf(\frac{\vert Z \vert}{\sqrt 2}, \infty)
= \frac{2}{\sqrt{2\pi}}\int_{\vert Z \vert}^{\infty} exp\left(\frac{-Z^2}{2}\right)dZ
\label{lh}
\end{equation}
where $Erf$ is the error function. 
For samples drawn from a normal distribution with unit standard
deviation, the values of $LH$ are evenly distributed between $0$ and
$1$ and the median value is about $0.5$ \citep{sch1}. Ideally, the normalized
residual should be normally distributed with zero mean and unit 
standard distribution. Hence, central tendency measures (Mean $Z$ 
and Median $Z$) close to zero indicate that the equation is unbiased
and the standard deviation of the residual (Std $Z$) close to $1$
indicates that the standard deviation of the GMPE adequately captures
that of the recorded data. A median value of the likelihood parameter
(Med. $LH$) indicates that the GMPE matches the data in terms of both
mean and standard deviation \citep{aran1}.

Based on the values of Mean $Z$, Median $Z$, Std $Z$ and Med. $LH$,
\citet{sch1} defined four categories of ranking of GMPEs. The values 
of the parameters for deciding the rank of a GMPE are given in
Table \ref{rank}.
\begin{table}
\caption{Values of different parameters for deciding the rank of a GMPE}
\label{rank}       
\begin{tabular}{ccccc}
\hline\noalign{\smallskip}
 Mean $Z$ & Median $Z$ & Std $Z$ & Med. $LH$ & Rank \\
\noalign{\smallskip}\hline\noalign{\smallskip}
$<$ 0.25 & $<$ 0.25  & $<$ 1.125  & $>$ 0.4 & A (high capability)  \\
$<$ 0.50 & $<$ 0.50  & $<$ 1.250  & $>$ 0.3 & B (medium capability)  \\
$<$ 0.75 & $<$ 0.75  & $<$ 1.500  & $>$ 0.2 & C (low capability)  \\
 & & All other combinations & & \\
 & & of parameters & & D (unacceptable capability)  \\
\noalign{\smallskip}\hline
\end{tabular}
\end{table}

The efficacy test was performed for a number of GMPEs corresponding
to recorded earthquake in each Layer. Owing to limited available 
strong-motion data, $85$ three-component strong-motion 
accelerograms (A total of $190$ horizontal accelerograms) 
from $8$ earthquakes of different hypo-central
depths have been considered for this purpose. Out of these
$190$ accelerograms, $44$ accelerograms belong to Layer $1$,
$72$ accelerograms belong to Layer $2$ and $74$ accelerograms
belong to Layer $3$. The strong-motion data, used for performing 
the efficacy test of GMPEs, have been 
downloaded from Center for Engineering Strong Motion Data 
(\url {https://www.strongmotioncenter.org}).
The details of the earthquakes considered are 
summarized in Table \ref{smeqs}.
\begin{table}[h]
\begin{center}
\caption{Details of the earthquakes considered for performing the efficacy test}
\label{smeqs}       
\begin{tabular}{lccclcc}
\hline\noalign{\smallskip}
	Name of & Date & Lat ($\degree N$) & Lon ($\degree E$) & $M$ & $H$ & No. of \\
 earthquake & (dd/mm/yy) & & & & (km) & records \\
\noalign{\smallskip}\hline\noalign{\smallskip}
	$\underline{Layer1}$ & & & & & & \\
 1. Sikkim & 18/09/2011  & 27.723 & 88.064 & 6.8 $M_W$  & 10 & 5\\
 2. Indo-Bangladesh  & 06/02/1988 & 24.688 & 91.570 & 5.8 $M_S$ & 15 & 17 \\
\noalign{\smallskip}\hline\noalign{\smallskip}
	$\underline{Layer2}$ & & & & & & \\
	3. Indo - Bangladesh  & 08/05/1997 & 24.894  & 92.250 & 6.0 $M_W$ & 34 & 11\\
	4. Indo - Bangladesh & 10/09/1986 & 23.385 & 92.077 & 4.5 $M_S$ & 43 & 12 \\
	5. Indo - Burma & 18/05/1987 & 25.271 & 94.202 & 5.9 $M_S$ & 49 & 13 \\
\noalign{\smallskip}\hline\noalign{\smallskip}
	$\underline{Layer3}$ & & & & & & \\
	6. Indo - Burma & 06/08/1988 & 25.149 & 95.127 & 7.2 $M_S$ & 90 & 17 \\
	7. Indo - Burma & 06/05/1995 & 24.987 & 95.127 & 6.4 $M_W$ & 117 & 09 \\
	8. Indo - Burma & 09/01/1990 & 24.713 & 95.240 & 6.1 $M_W$ & 119 & 11 \\
\noalign{\smallskip}\hline
\end{tabular}
\end{center}
\end{table}
Based on the suitability test, discussed above, regional and global
GMPEs have been selected for the estimation of seismic hazard of the 
region. The selected GMPEs with the references and codes in the brackets
are given in Table \ref{selgmpe}

\begin{figure}[!h]
\begin{center}
\begin{tabular}{cc}
      \resizebox{60mm}{!}{\rotatebox{0}{\includegraphics[scale=0.6]{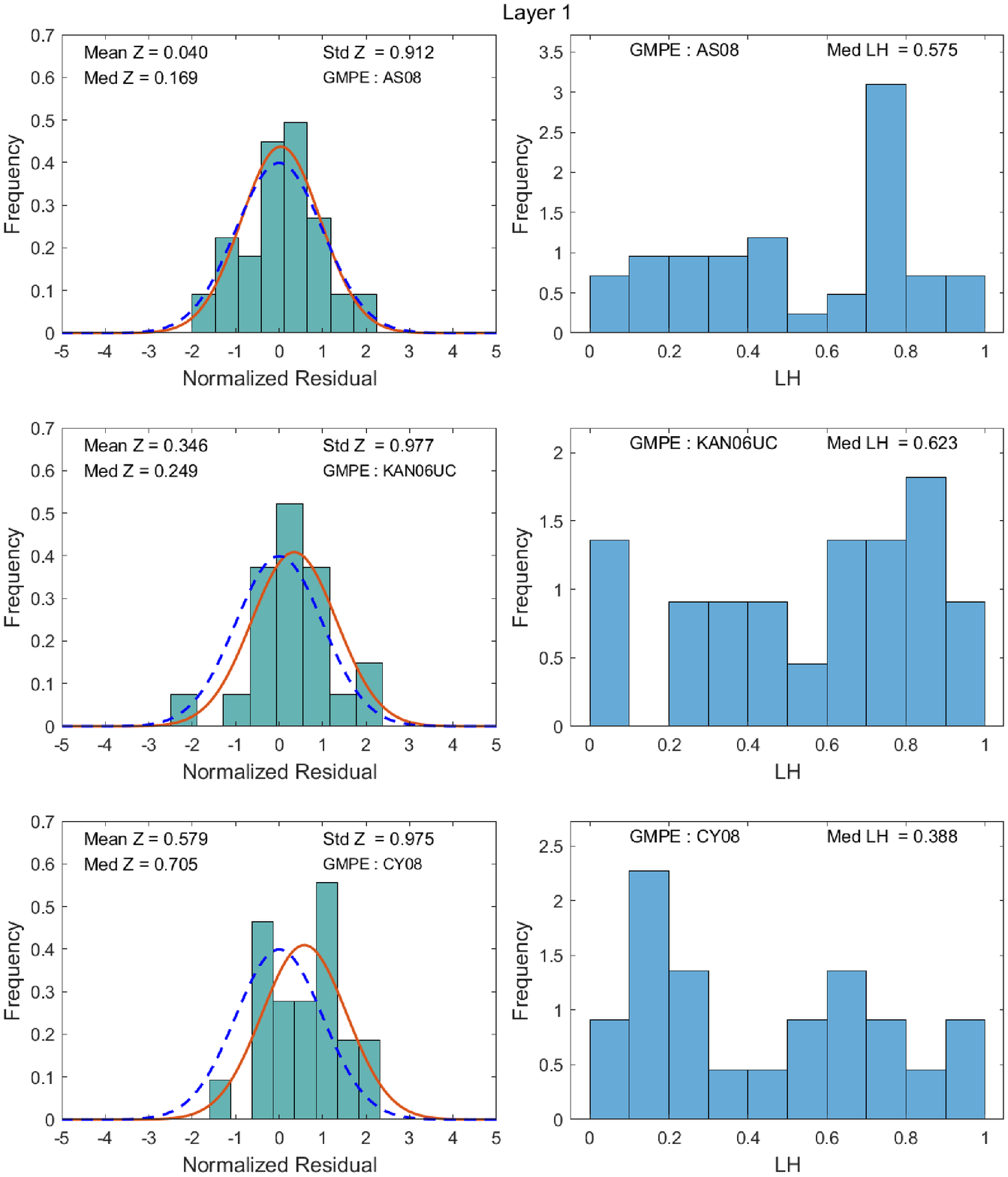}}} &
      \resizebox{60mm}{!}{\rotatebox{0}{\includegraphics[scale=0.6]{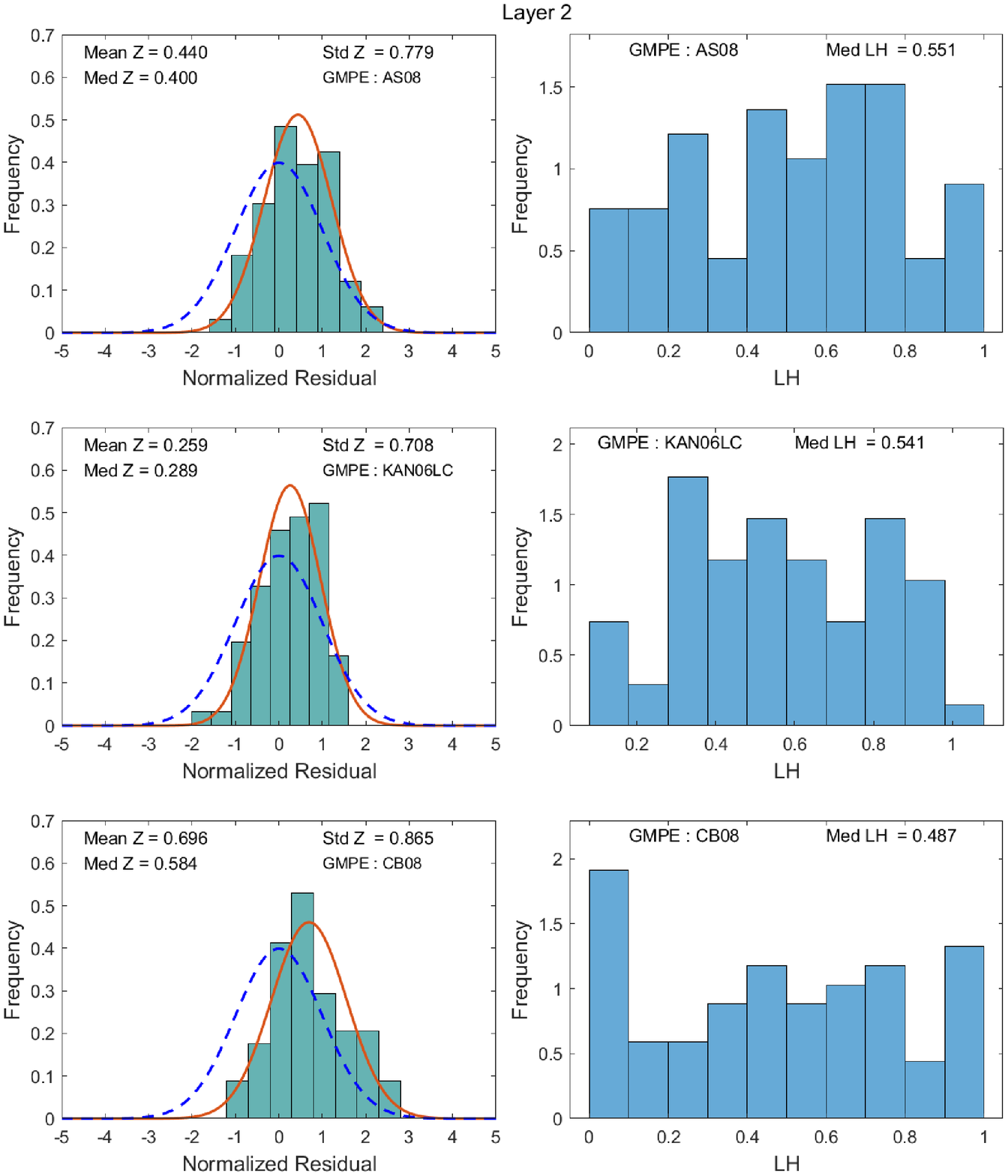}}} \\
      \resizebox{60mm}{!}{\rotatebox{0}{\includegraphics[scale=0.6]{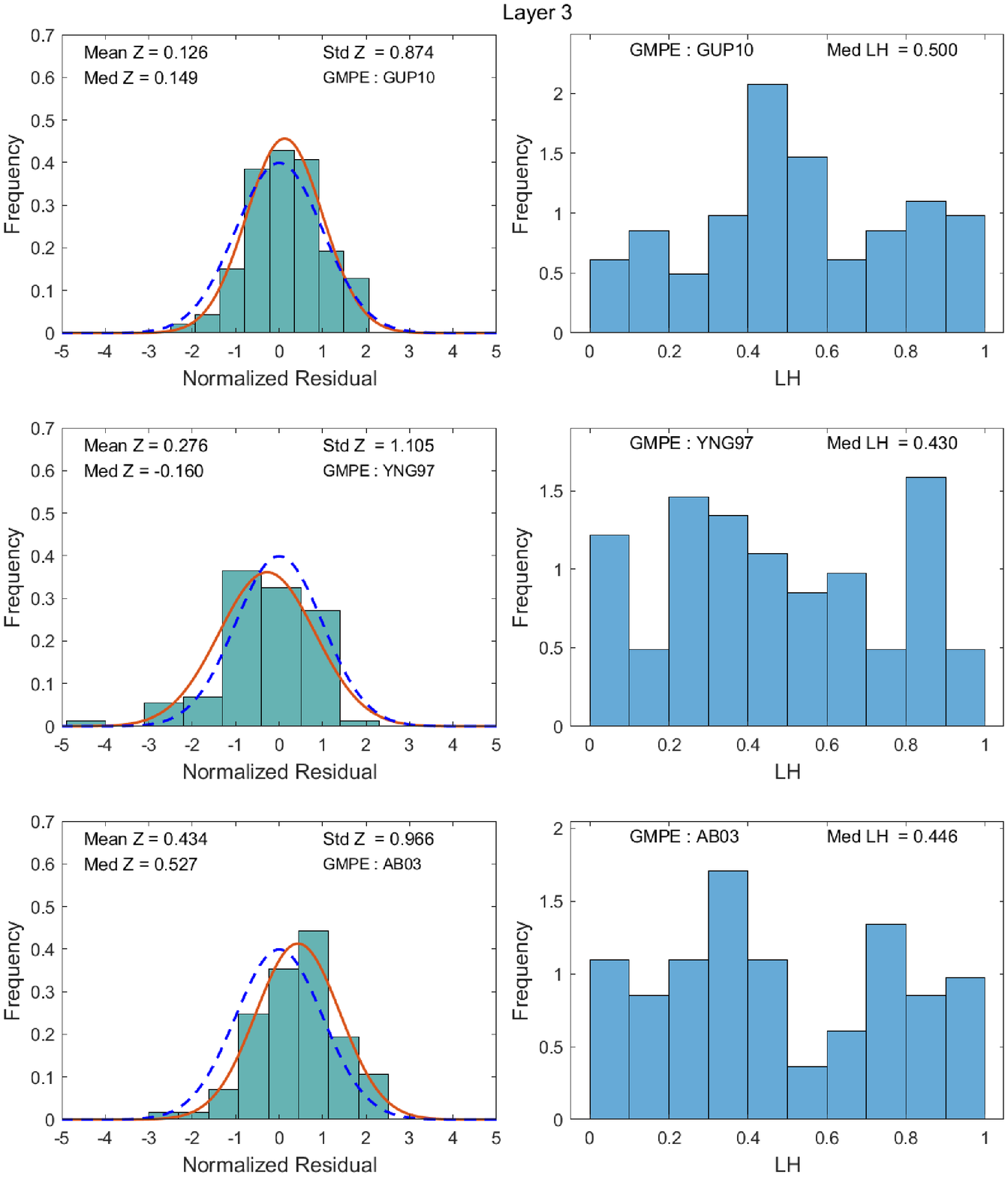}}} &
    \end{tabular}
\end{center}
\caption{Left side of each sub figure: Comparison of the distributions of normalized 
	residuals and the best-fit
	normal distributions (continuous curve) with the standard normal distribution (dashed
	curves) for the selected GMPEs with the mean, median and standard deviation of the 
	normalized residual indicated for all the three Layers. Right side of each sub figure: 
	Corresponding 
	histograms of the likelihood values with the median of LH indicated for all the three
	Layers.}
\label{lhfig}
\end{figure}

The comparisons of the histograms of the normalized residual
values (as obtained from Eqn. \ref{nr}) and the corresponding
approximation by the normal distribution (continuous curve) 
along with the standard normal distribution (dashed curve)
for three most suitable GMPEs for each Layer are plotted in
the left sides of each sub-figure of Figure \ref{lhfig}.  
The mean, median and standard deviation of the normalized residual ($Z$) 
are also indicated for each case. The corresponding right sides of each
sub-figure of 
Figure \ref{lhfig} plot the histograms of the $LH$ values (as obtained
from Eqn. \ref{lh}), with the corresponding median values indicated for each 
case. Based on the quantitative assessment, the ranking of the selected
GMPEs has been decided.
After that the weight factor to each GMPE for each Layer
has been decided. The higher the ranking of a GMPE, the more is the
weight factor. Therefore, from the set of GMPEs selected for each
Layer, the highest ranking GMPE has been assigned more weight and the
lowest ranking GMPE has been assigned less weight for the implementation
of logic tree approach. 
The ranking of the selected GMPEs and the corresponding weight factor 
for each layer are summarized in Table \ref{selgmpe}. 
For computations of seismic 
hazard from the deep-focused earthquakes in Layer $3$, the GMPEs used in 
Layer $2$ have been employed with the same weight factors as that of
Layer $2$.
\begin{table}[h]
\caption{Selected GMPEs used in the study along with their ranking
	and weight factor}
\label{selgmpe}       
\begin{tabular}{clcc}
\hline\noalign{\smallskip}
	& Name GMPEs with reference(s) & Rank & Weight factor   \\
\noalign{\smallskip}\hline\noalign{\smallskip}
	& \citet{abra2} (AS08) & A & 0.5 \\
	Layer 1 &  \citet{kann1} (KAN06UC) & B & 0.3 \\
	& \citet{chi1} (CY08) & C & 0.2 \\
\noalign{\smallskip}\hline\noalign{\smallskip}
	& \citet{abra2} (AS08) & B & 0.4 \\ 
	Layer 2 &  \citet{kann1} (KAN06LC) & B & 0.4 \\
	& \citet{cam4} (CB08) & C & 0.2 \\ 
\noalign{\smallskip}\hline\noalign{\smallskip}
	& \cite{gupta3} (GUP10) & A & 0.5 \\
	Layer 3 &  \citet{youn1} (YNG97) & B & 0.3 \\
	(subduction) & \citet{atkin1} (AB03) & C & 0.2 \\
\noalign{\smallskip}\hline
\end{tabular}
\end{table}

\subsection{Computation of Hazard}
The seismic hazard at a particular site is usually quantified in terms
of level of ground motion. Seismic hazard can be obtained for 
individual SSZ and then combined to express the aggrerate hazard at a
particular site. The methodology for PSHA incorporates how often 
annual rate of ground motion exceeds a specific value for different 
return periods of the hazard at a particular site of interest. The
probability of exceeding a particular value of $y^*$ of a ground motion
parameter $Y$ is calculated for one possible earthquake (of a particular
magnitude) at one possible source locations (source-to-site distance)
and then multiplied by the probability that the earthquake of that
particular magnitude would occur at that particular location. The process
is then repeated for all possible magnitudes and locations with the 
probabilities of each summed.

For a given earthquake occurrence, the probability that a ground motion
parameter $Y$ will exceed a particular value $y^*$ can be computed using
the total probability theorem \citep{kra1}
\begin{equation}
P\left[Y>y^*\right]=P\left[Y>y^*\vert \bf{X}\right]P\left[\bf{X}\right]
=\int P\left[Y>y^*\vert \bf{X}\right] f_x\left(\bf{X}\right)dx
\label{tph}
\end{equation}
where $\bf{X}$ is a vector of random variables that influences $Y$. For
the purpose of computing seismic hazard, the quantities in $\bf{X}$ are
limited to magnitude ($M$) and distance ($R$). Assuming that $M$ and $R$
are independent, the probability of exceedance can be written as 
\begin{equation}
P\left[Y>y^*\right]=\iint P\left[Y>y^*\vert m,r\right]f_M(m)f_R(r)~dm~dr
\label{poe1}
\end{equation}
where $P\left[Y>y^*\vert m,r\right]$ is obtained from the predictive
relationship and $f_M(m)$ and $f_R(r)$ are the probability density
functions (pdf) for $M$ and $R$ respectively.

If the site of interest is in a region of $N_s$ SSZs, each of which has
an average rate of threshold magnitude exceedance ( or annual activity
rate) $\lambda_i$, the total average exceedance rate for the region is
given by 
\begin{equation}
\nu_{y^*}=\sum_{i}^{N_s}\lambda_i \iint P\left[Y>y^*\vert m,r\right]
	f_{Mi}(m)\,f_{Ri}(r)\,dm\,dr
\label{aer1}
\end{equation}
The integral \ref{aer1} cannot be evaluated analytically for virtually 
all realistic PSHAs. Numerical integration is therefore required.
The approach, generally employed, is to discretize the possible 
ranges of magnitude and distance into $N_M$ and $N_R$ segments,
respectively. The average exceedance rate can then be estimated by
\begin{equation}
	\nu_{y^*} = \sum_{i=1}^{N_s} \sum_{j=1}^{N_M} \sum_{k=1}^{N_R}
	\lambda_i P\left[Y>y^*\vert m_j,r_k\right] f_{Mi}(m_j)\,
	f_{Ri}(r_k)\,\Delta m\,\Delta r
\label{aer2}
\end{equation}
This is equivalent to assuming that each SSZ is capable of generating 
$N_M$ different earthquakes of magnitude $m_j$ at $N_R$ different
source-to-site distances $r_k$. Eqn. \ref{aer2} is then equivalent to
\begin{equation}
	\nu_{y^*} = \sum_{i=1}^{N_s} \sum_{j=1}^{N_M} \sum_{k=1}^{N_R}
	\lambda_i P\left[Y>y^*\vert m_j,r_k\right] 
	P \left[M=m_j \right] P \left[R=r_k \right]
\label{aer3}
\end{equation}
The reciprocal of $\nu_{y^*}$ gives the return period for the ground
motion parameters $y^*$. The probability of exceedance of $y^*$ in
finite time intervals (say, for an exposure period $T$) can be
estimated from the Poisson model
\begin{equation}
P \left[Y_T > y^* \right]=1 - e^{-\lambda_{y^*}T}
\label{pom}
\end{equation}
The foregoing computational technique for PSHA is implemented for all
the grid points (sites) at engineering bedrock level ($V_{S30} \sim 
760\, m/s$) for $2\%$ and $10\%$ probabilities of exceedance in an 
exposure period of $50$ years. This corresponds to return periods
of $475$ years and $2475$ years respectively.

The different distance parameters like rupture distance ($R_{rup}$), 
horizontal distance to top edge of the rupture ($R_x$) etc, 
required for implementing the GMPEs, have been determined by using
the framework presented by \citep{kakl1} for estimating unknown 
input parameters.

\begin{figure}[h]
\begin{center}
	\rotatebox{0}{\includegraphics[scale=0.6]{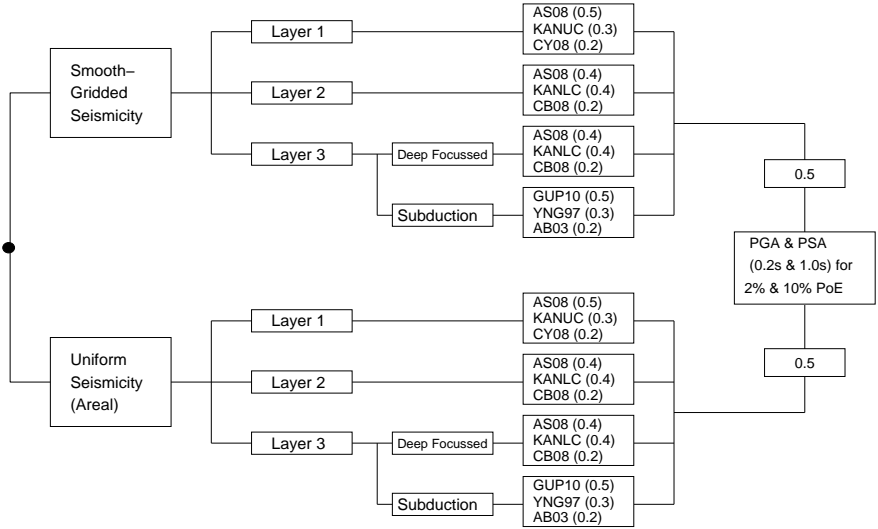}}
\caption{The logic tree framework employed in the present study}
\label{lt}       
\end{center}
\end{figure}

A logic tree framework approach has been adopted for the computation
of hazard at each site to incorporate multiple models in 
source considerations and GMPEs. Figure \ref{lt} depicts a logic
tree formulation at a site. In the present study, since the 
seismogenic source framework has been formulated with two
schemes, namely, smooth-gridded seismicity and uniform-seismicity
areal zones, both has been collectively assigned weight factor
equal to $0.5$. The hazard distributions are computed for the
SSZs at each depth-section separately and thereafter added.

\section{Result and Discussion}
\label{rd}
The hazard distributions are estimated for the seismogenic source
zones at three Layers corresponding to hypo-central depth ranges
of 0 - 25 km, 25 - 70 km and 70 - 180 km separately and added
thereafter. The seismogenic source zones include conventional
area sources of spatially uniform seismicity (uniform seismicity
areal zones) as well as spatially non-uniform seismicity (smooth-
gridded seismicity). The results obtained from both the uniform
and the non-uniform seismicity are integrated with equal weight
factor 0.5 to establish the overall hazard distribution in 
Tripura State at engineering bedrock level. The peak ground
acceleration (PGA) and 5$\%$ damped pseudo spectral 
acceleration (PSA) at 0.2 s and 1.0 s for all the sites
(grid points) for 10$\%$ and 2$\%$ probabilities of
exceedance (PoE) in an exposure period of 50 years (corresponding
to return periods of 475 years and 2475 years) have been computed
at engineering bedrock level conforming to $V_{S30} = 760$ m/s. 
The return period
of 2475 years represent the maximum considered earthquake (MCE)
condition while the return period of 475 years represents the 
design basis earthquake (DBE) condition \citep{ASCE,ICC}. PSA at 0.2 s
and 1.0 s are selected because those are frequently used as 
corner spectral periods to construct a smooth design spectrum for
structural designi \citep{nath4}.

\begin{figure}[!h]
\begin{center}
\begin{tabular}{cc}
      \resizebox{50mm}{!}{\rotatebox{0}{\includegraphics[scale=0.6]{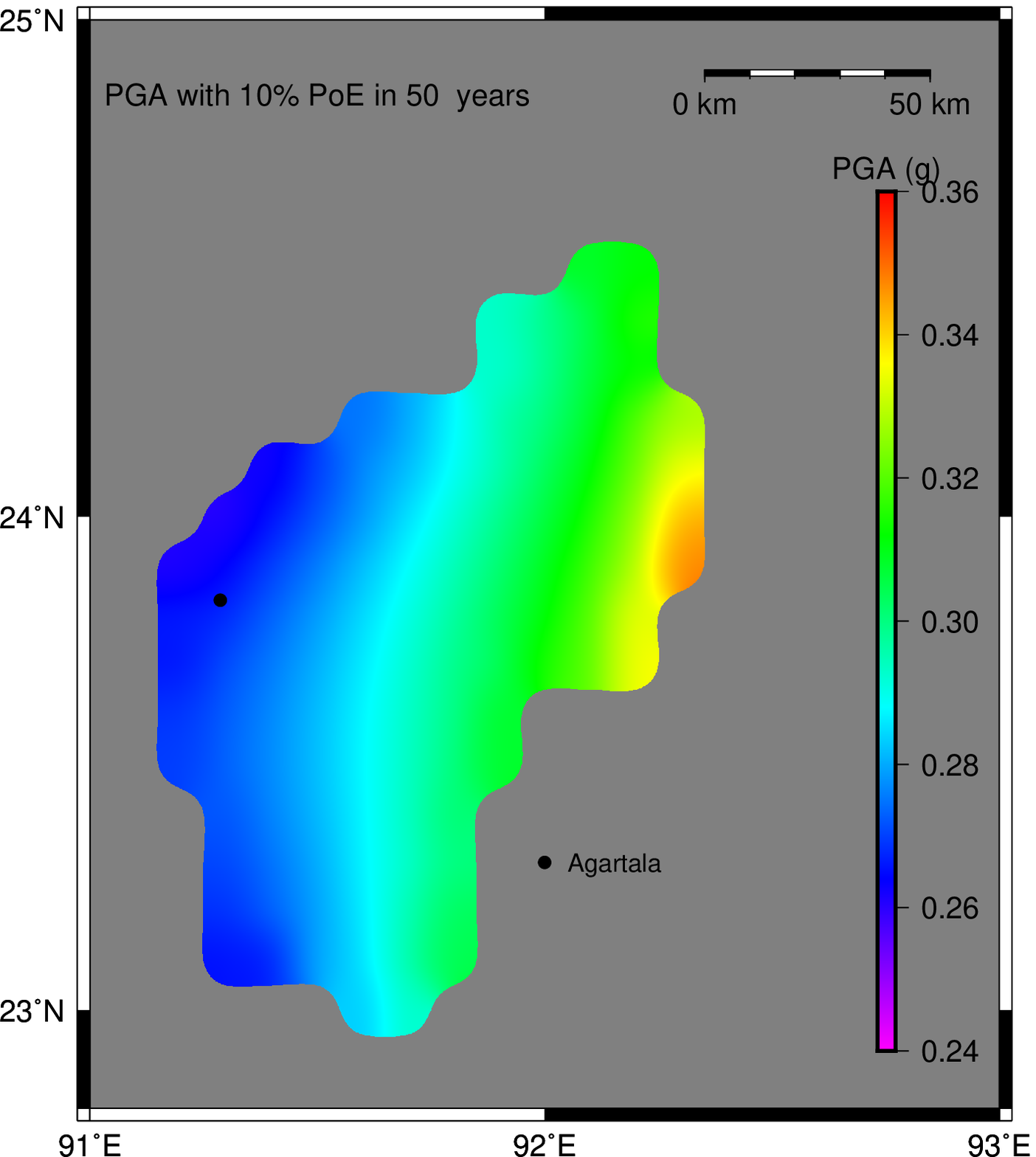}}} &
      \resizebox{50mm}{!}{\rotatebox{0}{\includegraphics[scale=0.6]{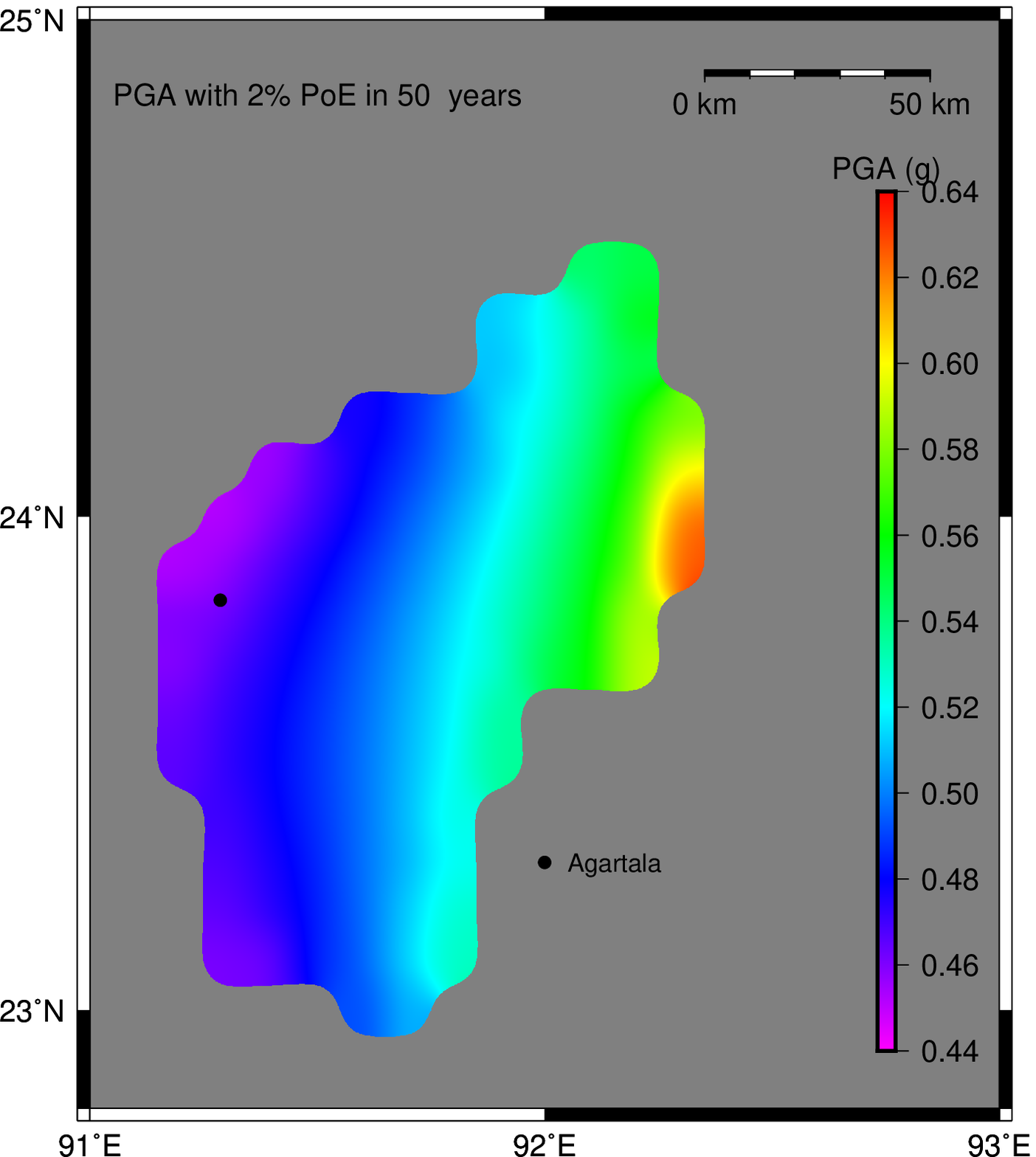}}} \\
      \resizebox{50mm}{!}{\rotatebox{0}{\includegraphics[scale=0.6]{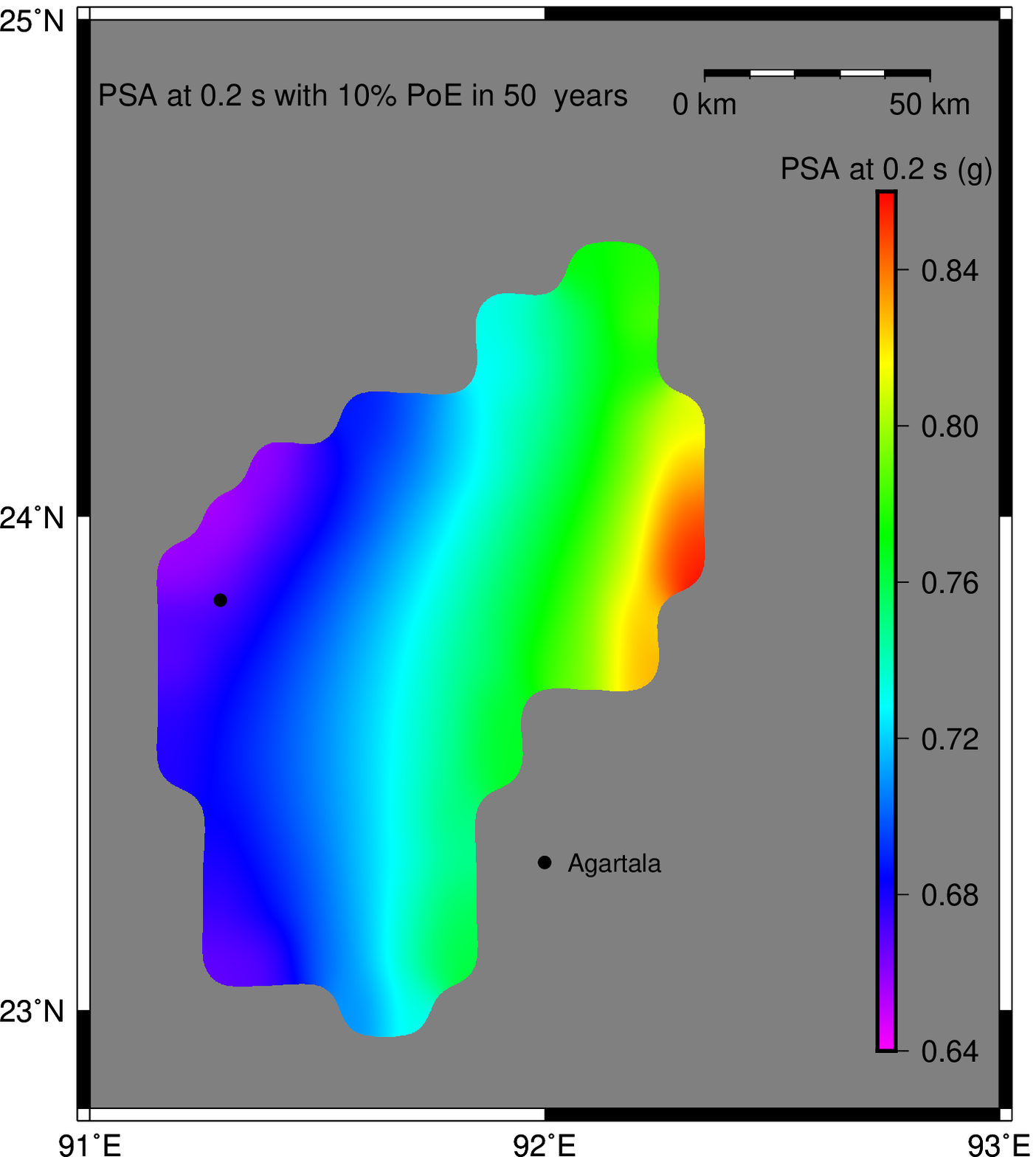}}} &
      \resizebox{50mm}{!}{\rotatebox{0}{\includegraphics[scale=0.6]{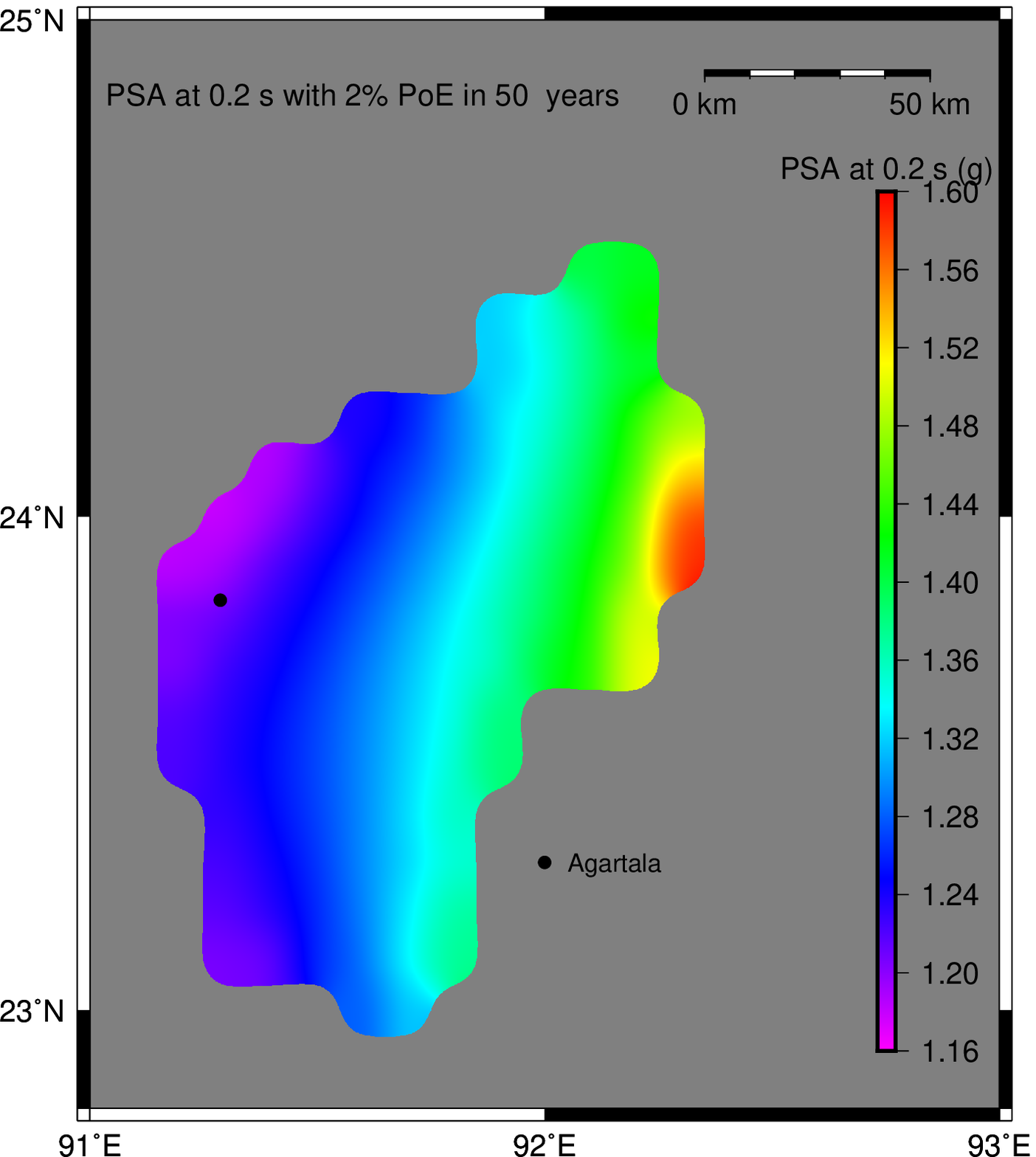}}} \\
      \resizebox{50mm}{!}{\rotatebox{0}{\includegraphics[scale=0.6]{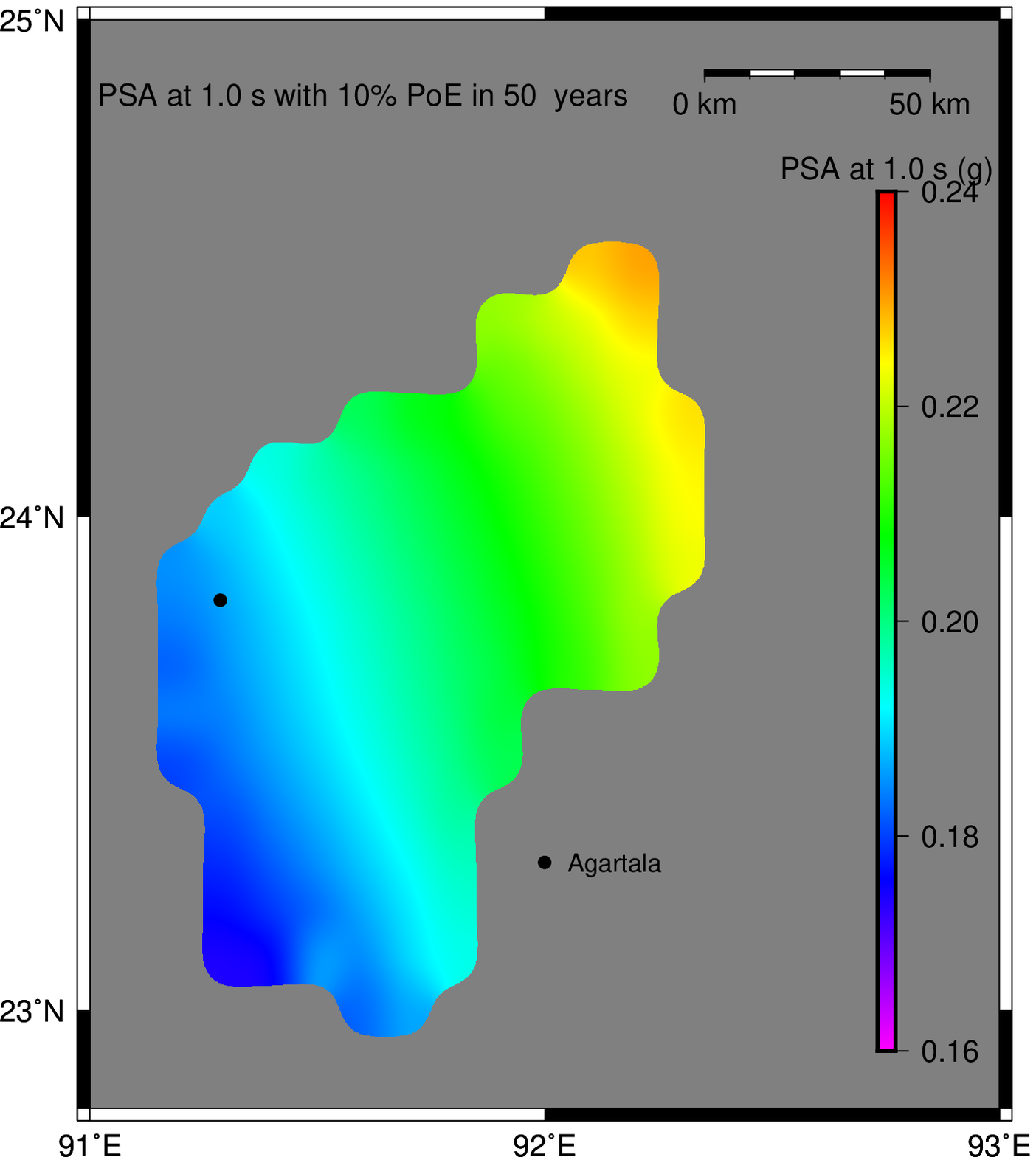}}} &
      \resizebox{50mm}{!}{\rotatebox{0}{\includegraphics[scale=0.6]{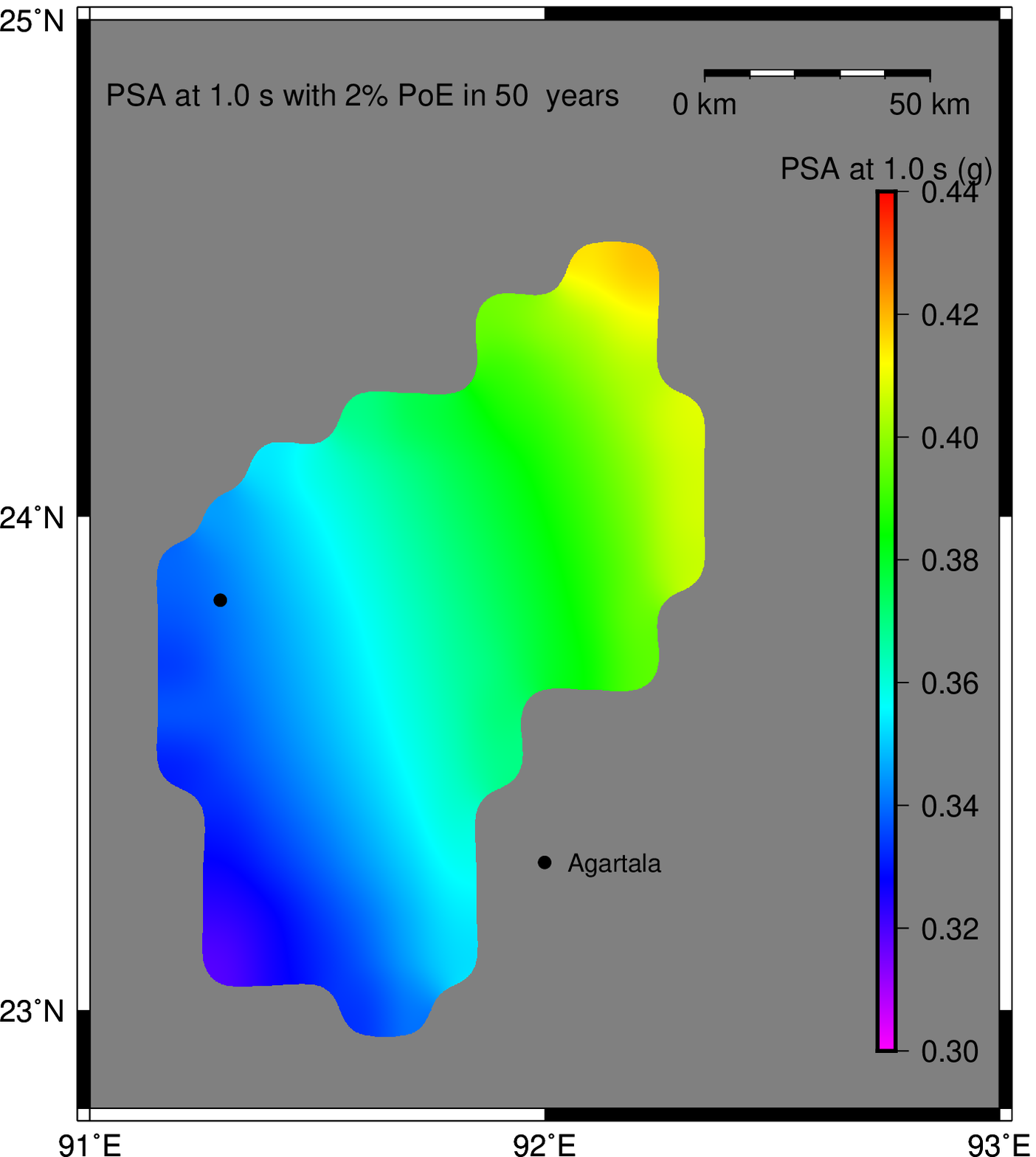}}} \\
    \end{tabular}
\end{center}
\caption{Seismic hazard distribution map of Tripura in terms of PGA and PSA at
	0.2 s and 1.0 s for 10$\%$ PoE in 50 years (left) and 2$\%$ PoE in 50
	years (right) at firm rock site conditions.}
\label{shzd}
\end{figure}

The seismic hazard maps in terms of the spatial distribution of
PGA and PSA at 0.2 s and 1.0 s, respectively for both DBE and MCE 
conditions at engineering bedrock is presented in Figure \ref{shzd}. The
left side plots in Fig. \ref{shzd} shows the hazard maps for a return
period of 475 years while the right-side plots shows the same for
a return period of 2475 years. For design purpose, 10$\%$ PoE in 
50 years are considered to be more appropriate and ideally used.

The PGA distribution for 10$\%$ PoE in 50 years at firm rock site shows 
variation from 0.2602 g to 0.3461 g for Tripura. The State capital 
Agartala has a hazard level to the tune of 0.264 g. The PSA at 0.2 s
exhibits a spatial variation between 0.6546 g to 0.8531 g while for
1.0 s, it ranges from 0.1745 g to 0.2298 g. The present seismic 
hazard analysis represents significant improvements over the 
deterministic zonation of \citet{bis1} and captures the local 
variation in seismic hazard well whereas \citet{bis1} suggested to
consider an uniform hazard value. The treatment to seismicity data
in different hypo-central depth ranges together with improved 
seismogenic source zonations and logic tree approach has produced
seismic hazard level different from those reported in earlier
studies by other researchers \citep{bha1,bis1,shar2,ndma,nath2,das2}. 
A comparison of the computed PGA values for 10$\%$ PoE in 50 years
with other studies at the State capital Agartala has been shown
in Table \ref{comp}
\begin{table}[h]
	\caption{Comparison of computed PGA with other studies for 10$\%$
	PoE in 50 years at Agartala city}
\label{comp}       
\begin{tabular}{cc}
\hline\noalign{\smallskip}
	Name of the studies & DBE level PGA   \\
\noalign{\smallskip}\hline\noalign{\smallskip}
	\citet{bha1}  & 0.45 \\
	\citet{bis1}  & 0.18 \\
	\citet{shar2} & 0.30 \\
	\citet{ndma} & 0.18 \\
	\citet{nath2} & 0.50 \\
	\citet{das2} & 0.217 \\
	Present Study & 0.264 \\
\noalign{\smallskip}\hline
\end{tabular}
\end{table}

The variation in PGA distribution for 2$\%$ PoE in 50 years
ranges from 0.4508 g to 0.6234 g while PSA at 0.2 s and 1.0 s
shows a variation of 1.1780 g to 1.5785 g and 0.3196 g to
0.4184 g, respectively. 
In the present study, north, east, northeastern and southeastern
parts of Tripura have
been found to show high hazard compared to other parts of the
State. The hazard maps obtained from the study is also able to
exhibit a significant local variation in seismic hazard.

\section{Conclusion}
\label{conclu}
The aim of the present study was to obtain the probabilistic
seismic hazard map of Tripura in engineering bed rock level
in terms of PGA and 5$\%$ damped PSA at 0.2 s and 1.0 s for
return periods of 475 years and 2475 years. An updated and
comprehensive earthquake catalogue has been employed to 
prepare the hazard maps. The catalogue is then first
homogenized into an unified moment magnitude and then 
declustered to remove the aftershocks and foreshocks. In
our opinion, the final results presented in the study show
significant improvements over the earlier studies. This can
be attributed to several factors - (a) A layered seismogenic
source framework and smooth-gridded seismicity models 
conforming to the variation of seismo-tectonic attributes
with hypo-central depth have been considered, (b) multiple
GMPEs, selected after performing a thorough quantitative assessment
with the recorded strong motion data, have been employed, thus
accounting for the epistemic uncertainties,
(c) the computations of hazard have been carried out in
a finer resolution of grid intervals and (d) the earthquake
database has been updated up to the year 2020. The seismic 
hazard maps prepared at the engineering bedrock level 
exhibit a significant spatial variation which is
consistent with the trends of the major tectonic features
and the past seismicity as well. The minimum and maximum
PGA for 10$\%$ PoE in 50 years (corresponding to return
period of 475 years) are found to be 0.2602 g and 0.3461 g,
respectively. The same for 2$\%$ PoE in 50 years 
(corresponding to return period of 2475 years) are 
reported as 0.4508 g and 0.6234 g, respectively.
As the seismic hazard
analysis is considered to be an integral part of the
earthquake-induced disaster mitigation practices, we 
hope that the present study is important towards 
updating the regional building code provisions for
earthquake-resistant design and construction of structures
in this high seismicity region.

\begin{acknowledgements}
One of the authors (S. Sinha) sincerely thanks Prof. A. Kijko for
providing the computer program for determining $M_{\tt max}$. 
Figures \ref{tecmap}, \ref{seistec}, \ref{ssz}, \ref{sgs},
\ref{shzd} were prepared by using the Generic Mapping Tools
(GMT) software package \citep{wess}, available at 
\url{https://www.generic-mapping-tools.org/}. The authors are also
thankful to Shri Rizwan Ali, Scientist - E and Shri A. K. Agrawal,
Director for continuous motivation towards conducting the
present research.
\end{acknowledgements}

\section*{Declaration of Conflicts of Interest and Funding}
The authors have no conflicts of interest to declare that
are relevant to the content of this article.
No funding was received for conducting this study.

\bibliographystyle{spbasic}
\bibliography{bibliography}

\end{document}